\documentclass[%
 reprint,
 nofootinbib,
 amsmath,
 amssymb,
 aps,
 prl,
]{revtex4-2}

\usepackage{persomain}

\usepackage{titlesec}

\titleformat{\section}[block]
  {\bfseries\centering} 
  {\thesection} 
  {1em} 
  {\MakeUppercase} 

\titleformat{\subsection}[hang]
  {\bfseries}
  {\thesubsection}
  {0.5em}
  {}

\titlespacing*{\section}
  {0pt}
  {12pt}
  {6pt}

\titlespacing*{\subsection}
  {0pt}
  {8pt}
  {0pt}

\begin{document}

\title{
	Coupled imbibition and evaporation of droplets deposited on a nanoporous layer
}

\author{Joachim Trosseille}
\author{Hugo Bellezza}
\author{Olivier Vincent}
\email{olivier.vincent@cnrs.fr}
\affiliation{
	Universite Claude Bernard Lyon 1, CNRS, Institut Lumi\`ere Mati\`ere,
	UMR5306, F-69100, Villeurbanne, France
}


\begin{abstract}

	Liquids in nanoscale hydrophilic pores generate capillary pressures so large that they could theoretically climb several kilometers against gravity.
	But droplets deposited on thin nanoporous layers form imbibition fronts that stop at only millimeters or less, due to evaporation competing with the capillary flow.
	Recently, there has been growing interest in such droplet infiltration dynamics, either as a way to study the behavior of confined fluids or in connection with applications, e.g., in water harvesting, printing, chemical delivery, actuation, sensing, etc.
	Here, we investigate both theoretically and experimentally the spontaneous imbibition and evaporation of sessile droplets into a thin mesoporous layer, focusing on their dependence on imposed relative humidity (RH).
	Theoretically, we provide a unified analytical approach for the dynamics of the wetted annulus (”halo”) that forms around the droplet.
	This approach accounts for halos of arbitrary dimensions and incorporates confinement-induced thermodynamic shifts (Kelvin effect).
	Experimentally, we study the case of water droplets deposited on oxidized porous silicon layers (pore diameter $\sim 3 \-- 4$, thickness $\simeq \qty{5}{\um}$) and systematically investigate how the halo and droplet dynamics depend on RH.
	In particular, we show that the time scales of halo formation diverge at a critical RH.
	This phenomenon is due to the Kelvin effect, which is clearly illustrated when comparing the dependence on RH of evaporation rates in the halo (confined liquid) and in the droplet (bulk liquid).
	Our analysis also shows an apparent divergence of the imbibition coefficient, which cannot be explained by standard capillary models.
	This observation suggests an important role of Kelvin-driven vapor transport along the porous surface.
	The complex couplings revealed by our study call for caution when interpreting halo dynamics data.
	Our results also demonstrate RH as a powerful control parameter for tuning droplet imbibition behavior and infiltration patterns.

\end{abstract}

\maketitle

\section{Introduction}

Liquids may spontaneously invade confined spaces due to surface tension effects (capillary action).
Capillary imbibition, or wicking, is an everyday life occurrence, e.g., when dipping sugar in coffee, cleaning with a sponge, or inserting a straw in a beverage.
But it also plays a role in a large variety of natural phenomena such as soil humidification or salinization \cite{Olivella2000,Shokri-Kuehni2017}, hydration of plants and wood \cite{Jensen2016,Ha2020,Zhou2019}, animal feeding and drinking \cite{Kim2012,Siefert2025}, etc.
In technology, textile performance, paper printing, drug delivery and a variety of microfluidic applications also take advantage of capillary imbibition \cite{Fischer2022,Venditti2021,Li2021,Olanrewaju2018}.
Conversely, wicking can be detrimental for heritage preservation, insulation materials in buildings, road engineering or food packaging \cite{Pel2018,Binder2010,Al-Samahiji2000,Todorova2022}.

These contexts and many others have inspired a broad scientific literature \cite{Masoodi2017}.
Historically, the description of imbibition dynamics dates back from the early $20\textsuperscript{th}$ century \cite{Bell1906}, in particular with the work of Lucas and Washburn \cite{Lucas1918,Washburn1921}.
The so-called Lucas-Washburn (LW) law predicts a penetration length proportional to $\sqrt{t}$ ($t$, time), due to a constant driving force (capillary pressure, $\DeltaPc \sim \gamma / \bar{r}$ from Laplace's law, with $\gamma$ the surface tension and $\bar{r}$ the radius of curvature of the liquid-vapor interface) and increasing viscous friction within a continuously growing wetted zone.

In the past decades, there has been increasing interest about capillary flows in channels or pores with dimensions in the nanoscale range \cite{vanHonschoten2010,Huber2015}.
At these scales, deviations from classical laws of capillary flows such as LW are observed, and have been attributed to changes in the boundary condition at the liquid/wall interface (slipping \cite{Dimitrov2007,Joly2011} or sticking, i.e. immobile layers at the wall \cite{Gruener2009,Vincent2016}), increasing importance of disjoining pressure and noncontinuum effects \cite{Gravelle2016}, confinement-induced modifications of physical parameters such as viscosity and surface tension \cite{Kelly2016} etc.
Nanoscale liquids also exhibit thermodynamic properties different from the bulk, e.g., shifted equilibrium vapor pressure due to the Kelvin effect \cite{Elliott2021}, which significantly impacts imbibition dynamics when phase change is involved \cite{Vincent2017,Vincent2024}.

Independently of these specific phenomena, capillary flows at the nanoscale generally share two common features.
First, because $\bar{r}$ is small, the magnitude of the capillary pressure, $\DeltaPc$ driving imbibition is exceptionally large, with typical magnitudes in the range of tens to hundreds of atmospheres.
These values allows capillary rise to theoretically reach hundreds of meters to several kilometers against gravity, according to Jurin's law \cite{Huber2015}.
Second, the hydraulic resistance and corresponding viscous dissipation are also extremely large (e.g., from Poiseuille's law, the conductivity of a cylindrical pore scales as its radius to the fourth power).
This low permeability to liquid flow makes capillary ascent particularly inefficient, and makes time scales to reach large heights unattainable in practice \cite{Caupin2008}.
Another consequence is that capillary flow is easily perturbed by competing transport mechanisms, such as evaporation.

In fact, evaporation is so efficient at limiting capillary imbibition in nanoporous materials that it can restrict capillary rise to below mm-cm distances instead of the theoretical rise of several kilometers \cite{Huber2015}.
This effect originates from evaporative loss of liquid from the material, which compensates capillary pumping and results in a dynamic arrest of the imbibition front.
Recently, this phenomenon has gained interest in the situation of imbibition around droplets deposited on thin porous media, motivated both by applications in advanced materials (e.g., for sensing, catalysis, actuation, water harvesting, energy conversion, etc.) and as a fundamental tool to study the dynamics of nanoconfined fluids in relation to pore structure \cite{Seker2008,Ceratti2015,Gimenez2018,Gimenez2018a,Urteaga2019,Khalil2020a,Hartmann2025}.
This situation is also relevant in the context of inkjet printing \cite{Venditti2021,Aslannejad2021}, spray-cooling techniques \cite{Weickgenannt2011} or damage from sea spray deposition in cultural heritage and building materials \cite{Chabas2000}.

In such geometries, imbibition results in an annular wetted zone of the porous material around the droplet, i.e., a \emph{halo}, which rapidly stops growing to reach typically sub-millimeter widths.
Several authors have proposed modified LW equations to describe halo dynamics \cite{Seker2008,Liu2016,Mercuri2017}.
In particular, the model proposed by Mercuri et al. \cite{Mercuri2017} predicts an exponential relaxation to a steady-state halo, which fits well experimental data and has been applied to more complex situations involving e.g. solutes \cite{Pizarro2024,Czerwenka2025}.
One of the key ingredient of these models is the evaporation rate from the porous medium, which is often difficult to predict and used as a fitting parameter.
Understanding and controlling evaporation rates is thus crucial when predicting halo dynamics and designing related applications.
Despite this importance, the impact of parameters such as the relative humidity (RH) in the environment has received little attention, while being one of the main driving forces for evaporation.

Here, we report on theoretical and experimental work about the dynamics of imbibition of droplets into nanoporous layers, and on its dependence on RH.
We provide a unified theory, which updates existing models to take into account flow coupling due to conservation of mass, and to incorporate the Kelvin effect.
We also propose a two-dimensional version of the model to describe cases where the halo size is non-negligible compared to the droplet radius.
Experimentally, we have investigated water droplets deposited on a hydrophilic, mesoporous layer (oxidized porous silicon, pore diameter $\simeq$ \qtyrange{3}{4}{\nm}) of thickness $\simeq \qty{5}{\um}$, for relative humidities in the range \qtyrange{5}{80}{\percent}RH.
As expected, RH has a strong, nonlinear impact on halo dynamics, with an apparent divergence of the halo size at a critical RH far below \qty{100}{\percent}RH.
We interpret this divergence as being due to the Kelvin effect, which we clearly demonstrate when comparing evaporation rates from the halo (confined water) and from the droplet (bulk liquid).
We also report an unexpected dependence on RH of the imbibition coefficient, which cannot be explained with capillary flow models such as LW.
We discuss at the end of the paper potential mechanisms that can explain this surprising dependence, and provide physical insight about their relative importance.

\section{Materials and methods}
\label{sec : methods}

\begin{figure}[]
	\centering
	\includegraphics[scale=0.9]{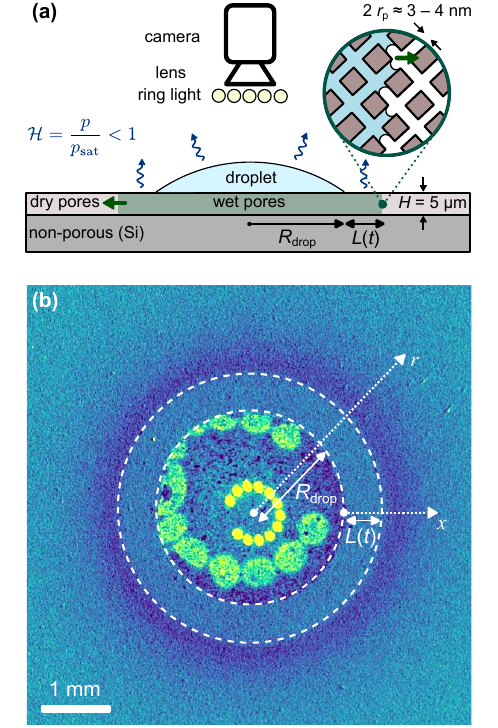}
    \caption{
		\small
		Experimental setup.
		(a) In a chamber controlled in relative humidity ($\rh$), we deposit a water droplet on a thin porous layer (oxidized porous silicon).
		Spontaneous imbibition results in a wetted zone in the pores (the \emph{halo}) of extension $L(t)$; $\idx{R}{drop}$ is the fixed radius of the pinned, sessile drop.
		The inset shows a schematic close-up view of the imbibition front at the edge of the halo.
		A camera with a macro lens and LED ring allows us to record top view images of the drop.
		(b) Pseudo-color, background-subtracted image of a \qty{1}{\micro\liter} droplet at a relative humidity of $\rh = 0.3$ (\qty{30}{\percent}RH), at a time, $t = \qty{53}{\second}$ after deposition on the sample.
		The inner dashed circle represents the droplet's contact line on the substrate, and the outer dashed circle shows the external limit of the halo.
		The bright circular shapes visible inside the drop are optical reflections of the LED illumination, which we use to estimate the droplet's shape and volume.
	}
	\label{fig : ExpSetup}
\end{figure}

Figure \ref{fig : ExpSetup}a presents our experimental approach.
We placed a thin layer of oxidized mesoporous silicon in a chamber with controlled RH and temperature.
On the porous surface, we deposited sessile water droplets, which invaded the pores due to capillarity, while simultaneously evaporating into the surrounding air.
Top-view images such as that in Figure \ref{fig : ExpSetup}b show the droplet (base radius, $\idx{R}{drop}$) and the annular, wetted zone of porous medium surrounding the droplet.
This zone is the so-called \emph{halo}, of width, $L(t)$.
With time-lapse imaging and image analysis, we recorded the temporal evolution of the halo dimensions, but also of the droplet shape, which we inferred from the reflected image of the ring light illumination visible in the center of the droplet.
Below we provide details on the various aspects of the experimental procedure.

\subsection{Sample fabrication}

We fabricated a layer of oxidized mesoporous silicon (poSi) of thickness, $H \simeq \qty{5}{\um}$, by etching the surface of a \qty{500}{\um}-thick silicon wafer ($\langle 111 \rangle$ crystal orientation, p-type, resistivity \qtyrange{1}{10}{\ohm\cm}), using hydrofluoric acid anodization.
The fabrication follows a procedure detailed elsewhere \cite{Vincent2014}.
We also oxidized the sample for 2 hours at \qty{800}{\degreeCelsius} under pure oxygen atmosphere in order to stabilize the surface and make it hydrophilic.
After this process, the porous layer at the surface became transparent, due to the silicon walls turning into silica.
The resulting porous layer contained connected, isotropic pores enabling lateral transport \cite{Vincent2016,Vincent2017}.
Eventually, we diced the wafer to obtain square pieces with lateral dimensions of $\ell \simeq 1$ cm.

\subsection{RH response and Kelvin effect}

We first studied the response of the porous layer (without deposited droplet) to quasi-static variations in RH, i.e., we measured water sorption isotherms.
The relative humidity is defined as
\begin{equation}
	\label{eq : rh}
	\rh = p / \psat
\end{equation}
where $p$ (\unit{\pascal}) is the partial vapor pressure of water in the air, and $\psat$ its saturation value.
The value of $\rh$ is thus comprised between 0 and 1.
For convenience, we often express it below as a percentage of relative humidity: in this context, \qty{60}{\percent}RH needs to be understood as $\rh = 0.6$.

\begin{figure}[]
	\centering
	\includegraphics[scale=0.9]{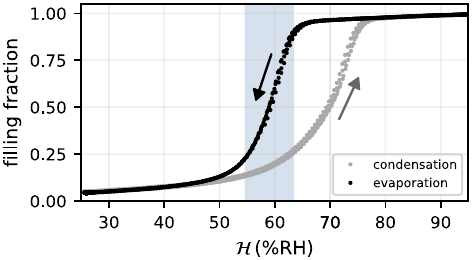}
    \caption{
		\small
		Water sorption isotherm: liquid filling fraction of the pores, $f$, as a function of the imposed relative humidity, $\rh = p / \psat$, following either condensation (increase of RH, gray dots) or evaporation (decrease of RH, black dots) branches.
		Five successive condensation/evaporation cycles are shown.
		The data was obtained with a method based on white light interferometry.
		The shaded, blue region corresponds to the typical equilibrium RH of the confined fluid, $\idx{\rh}{eq}$, see text.
	}
	\label{fig : Isotherm}
\end{figure}

We measured water sorption isotherms on the poSi sample using white light interferometry \cite{Casanova2012,Bossert2020} (WLI), using an experimental setup described elsewhere \cite{Bellezza2025}.
WLI consists in recording changes in the optical index of the sample, from which variations in the water content of the pores can be calculated using effective medium theory (see details in the \emph{Supporting Information}).
The resulting isotherms (Figure \ref{fig : Isotherm}) show a transition from a full state at high humidity (water filling fraction, $f \simeq 1$) to an empty state at low $\rh$ ($f$ approaching 0).

When lowering $\rh$, water spontaneously evaporated from the pores at a RH significantly below \qty{100}{\percent} (shaded blue area in Figure \ref{fig : Isotherm}).
This evaporation transition happens when the imposed RH reaches (and becomes lower than) the equilibrium RH of the confined liquid, $\idx{\rh}{eq}$.
Due to the so-called \emph{Kelvin effect}, $\idx{\rh}{eq} < 1$ is shifted from the bulk saturation RH ($\idx{\rh}{sat} = 1$), because of the curvature of the liquid-vapor interfaces in the pores, assumed to have spherical cap shapes.
This shift is more pronounced when the pores are smaller, as described by the Kelvin-Laplace equation \cite{Elliott2021,Bellezza2025}
\begin{equation}
	\label{eq : KelvinRH}
	\idx{\rh}{eq} = \exp \left( - \frac{2 \gamma \vm}{\rk \idealR T} \right)
\end{equation}
where $\rk$ is the radius of curvature (\unit{\m}) of the menisci at the evaporation point, which is on the order of the pore radius, $\idx{r}{p}$ (see \emph{Supporting Information}); the other parameters are $\gamma$ (\unit{\newton\per\meter}) the surface tension of the liquid-vapor interface $\vm$ the molar volume of water (\unit{\meter\cubed\per\mol}), $\psat$ the saturation vapor pressure of water (\unit{\pascal}), $\idealR$ the ideal gas constant (\unit{\joule\per\mole\per\kelvin}), $T$ the temperature (\unit{\kelvin}).
From the shaded blue area in Figure \ref{fig : Isotherm}, we estimate $\idx{\rh}{eq} = \qty{59.0(4.3)}{\percent}$RH.

Using Laplace's law, $\left( \DeltaPc \right)_\mathrm{eq} = - 2 \gamma / \rk$ is also the capillary pressure in the pore liquid, so that from Equation \ref{eq : KelvinRH},
\begin{equation}
	\label{eq : KelvinPc}
	\left( \DeltaPc \right)_\mathrm{eq} = \frac{\idealR T}{\vm} \ln (\idx{\rh}{eq})
\end{equation}
represents the equilibrium capillary pressure in the pores.
The evaporation point at $\idx{\rh}{eq}$ can thus also be understood as being due to the liquid pressure difference between the pore liquid and the external gas phase reaching $\left( \DeltaPc \right)_\mathrm{eq}$, causing meniscus recession and subsequent evaporation.
More generally, isotherms $f(\rh)$ such as in Figure \ref{fig : Isotherm} can be translated into pressure/saturation curves $\Delta P(f)$ using Equation \ref{eq : KelvinPc}.
As a result, any partially saturated zone, corresponding to the shaded blue area in Figure \ref{fig : Isotherm}, should contain liquid at a capillary pressure in the vicinity of $\left( \DeltaPc \right)_\mathrm{eq}$.

It is usually considered that the condensation branch of the isotherms (gray curve in Figure \ref{fig : Isotherm}) corresponds to metastable capillary condensation from adsorbed films (cylindrical menisci) in the pores when increasing RH \cite{Thommes2015}.
We estimate that capillary condensation occurs in our sample in a range $\idx{\rh}{cond} = \qty{68.4(5.9)}{\percent}$RH (see \emph{Supporting Information}).
In our experiments where liquid imbibition form hemispherical menisci in the pores (Figure \ref{fig : ExpSetup}), we assume that any local equilibrium should rather follow the evaporation branch rather than the condensation branch of the isotherms.

\subsection{Sample properties}

From the range $\idx{\rh}{eq} = \num{0.590(0.043)}$ and Equation \ref{eq : KelvinPc}, we calculated the equilibrium capillary pressure of water in the porous medium, $\left( \DeltaPc \right)_\mathrm{eq} = \qty{-72.7(9.9)}{\mega\pascal}$, corresponding to an approximate pore diameter of \qtyrange{3}{4}{\nano\meter} (see \emph{Supporting Information}).
We also extracted from the WLI data the thickness of the porous layer, $H = \qty{4.95(0.10)}{\um}$ and its porosity, $\phi = \num{0.39(0.02)}$ (see \emph{Supporting Information}).

\subsection{Experimental setup}

Our experimental system was enclosed in a semihermetic plastic box (dimensions of 10 cm $\times$ 10 cm, with a height of approximately 1.5 cm) with two lateral, symmetrical air inlets.
A copper plate containing a water circuit connected to a circulated thermostat (Bioblock) fixed the sample temperature.
A humid air generator from Instec (RHC01 with a MK2000 controller) provided an air flow of controlled relative humidity (accuracy $\pm$ \qty{1.8}{\percent}RH), using a mix between dry air and humidified air, at a rate of 3 L/min.
Given the box dimensions and the air flow, a typical air velocity associated with air renewal in the chamber is $v \simeq \qty{4}{\cm\per\second}$.
We recorded RH and temperature with a sensor (Sensirion SHT85, accuracy $\pm$ \qty{1.5}{\percent}RH and $\pm$ \qty{0.1}{\degreeCelsius}) attached to the copper plate in the vicinity of the sample.
A platinum probe (PRT Pt100) of accuracy $\pm$ \qty{0.1}{\degreeCelsius} also monitored the temperature of the air inside the box and around the sample.
For image acquisition, we used a 5MPx USB3 camera (Jai Go) fitted with a 50 mm, f/2.8 macro lens (Tamron) through a 40mm extension tube, resulting in a scale of $113.9$ pixels per mm.
The system was illuminated with a custom-made LED ring.

Since evaporation of a droplet is highly sensitive to humidity and temperature variations, we took particular experimental precautions to stabilize the sample temperature as close as possible to the room temperature (itself controlled with air conditioning).
Across all experiments, we obtained a stable sample temperature of \qty{24.3(0.4)}{\degreeCelsius}, and typical fluctuations of $\pm$ \qty{0.92}{\percent}RH of the relative humidity of the air.
Adding to these fluctuations the sensor precision (see above), we estimate a total uncertainty on relative humidity of $\pm$ \qty{2.5}{\percent}RH.

\subsection{Droplet deposition}

Once the temperature and RH were stabilized, we deposited a drop of volume \qty{1}{\micro\liter} in the center of the sample surface with a micro-pipette, taking particular care to obtain circular sessile drops
We performed this operation through a small hole in chamber wall, which we opened and closed quickly, so that the inside air humidity was perturbed as little as possible.
The deposition moment defined the time, $t = 0$ of each experiment.
We recorded images before and after deposition with the optical system described above, at an initial framerate of 2 fps later lowered to $0.1$ fps.

\subsection{Image analysis}

Figure \ref{fig : ExpSetup}b shows a typical image obtained after deposition of the droplet.
In order to enhance the visibility of the halo, we performed background subtraction, using an image prior to droplet deposition as reference.
In Figure \ref{fig : ExpSetup}b and in the rest of the paper, this processed image is displayed in pseudocolors, using the perceptually uniform colormap \emph{viridis}.
From such images, we extracted various information by image analysis (see \emph{Supporting Information} for details).

First, we determined the circular contact line of the drop with the substrate (Figure \ref{fig : ExpSetup}b, inner white dashed contour).
Across all our experiments, we measured a remarkably consistent radius of this contact line, $\idx{R}{drop}$ = \qty{1.34(0.04)}{\milli\meter}.

Second, we measured the limit of the imbibition front, i.e., the extension of the halo, of width $L(t)$ (Figure \ref{fig : ExpSetup}b, outer white dashed circle).
We only considered the fully saturated zone of the porous medium as being part of the halo (see \emph{Halo width and partially filled zone} below).

Finally, we used the ring of bright circular dots visible close to the center of the droplet in Figure \ref{fig : ExpSetup}b to estimate the droplet geometry.
Indeed, these dots correspond to primary optical reflections of the LED ring light on the top surface of the droplet, which acts as a spherical mirror.
From geometrical considerations, we thus calculated the droplet curvature, and deduced its contact angle, $\theta$, and volume, $V$ (see \emph{Supporting Information}).
The initial contact angle of the drop after deposition was also very consistent, with $\theta$ = \qty{27(2)}{\degree} for all experiments.
The larger and dimmer spots located closer to the droplet edge are secondary reflections of the ring light, which we have not used in our analysis.

\subsection{Halo width and partially filled zone}

The darker zone situated outside of the largest dashed circle in Figure \ref{fig : ExpSetup}b indicates increased light scattering due to partial pore filling \cite{Page1993,Jain2019}, resulting in radial pore filling gradients that have also been measured elsewhere with interferometric methods \cite{Sallese2020,Martinez2024}.
In our study, we only focus on the dynamics of the fully saturated zone (the halo area comprised between the two dashed circles in Figure \ref{fig : ExpSetup}b), for several reasons.
First, this wet zone is defined unambiguously from its uniform appearance in the images (see Figure \ref{fig : ExpSetup}b).
Second, we know from sorption isotherms that the pressure at the edge of this zone, $\DeltaPc$ should be in the vicinity of the equilibrium pore capillary pressure, $\left( \DeltaPc \right)_\mathrm{eq}$ (see \emph{RH response and Kelvin effect} above).
Finally, transport within that zone is single-phase (liquid) so that it can be readily modeled with Darcy's law, resulting in a Lucas-Washburn-like equation that is solvable analytically (see \emph{Theory} below).

\section{Theory}
\label{sec : theory}

We consider the situation described in Figure \ref{fig : ExpSetup}, where the droplet has a pinned contact line and constant base radius $\idx{R}{drop}$ and where a wetted halo, of lateral extent $L(t)$, develops in the porous medium due to capillary imbibition.
At the same time, the liquid within this halo evaporates into the air above the medium;
as the wetted area grows, the total evaporation rate from the halo increases, until reaching values comparable with the imbibition flow.
A steady-state situation is reached with a maximum halo extension, $\idx{L}{max}$, defined by the dynamic equilibrium between capillary flow and evaporation.
This steady-state persists until depinning and subsequent disappearance of the droplet, which we observe experimentally (see \emph{Results and Discussion}) but do not model here.

We derive below the equations that govern the dynamics of the halo, $L(t)$ during the expansion and steady-state phases.
First, we use a one-dimensional (1D) approximation and provide an equation similar to that developed by Mercuri et al. \cite{Mercuri2017}, with a significant correction imposed by mass conservation between evaporation and imbibition fluxes.
Second, we extend existing two-dimensional equations (2D) from the literature \cite{Hyvaluoma2006,Liu2016} to describe cases where the halo extension is non-negligible with respect to the size of the droplet.
We obtain similar analytical solutions in the 1D and 2D cases, with the introduction of effective quantities in the 2D case.
Finally, we also discuss how to estimate evaporation rates both from the droplet and from the halo (pore liquid), the latter being a key ingredient in predicting time scales of halo development in the theory.

\subsection{General considerations}

Spontaneous liquid imbibition in the pores occurs due to a pressure difference between the droplet ($P=P_0$) and the menisci at the advancing imbibition front ($P = P_0 + \DeltaPc < P_0$), where $\DeltaPc < 0$ is the capillary pressure at the front.
In our treatment, we neglect the capillary pressure associated with the macroscopic curvature of the droplet surface.
As a result, the boundary conditions of pressure in the halo are $P=P_0$ at $r = \idx{R}{drop}$ and $P = P_0 - \DeltaPc$ at $r = \idx{R}{drop} + L(t)$.

We assume that fluid flow in the pores responds to this pressure imbalance following Darcy's law, $\idx{q}{p} = - \rho \kappa \nabla P$, where $\rho$ is the liquid density (\unit{\kg\per\meter\cubed}) $\kappa$ is the Darcy permeability of the porous layer (\unit{\m\squared\per\pascal\per\second}), $\nabla P$ is the pressure gradient (\unit{\pascal\per\meter}), and $\idx{q}{p}$ is the mass flux (\unit{\kg\per\m\squared\per\second}), i.e., the mass flow rate of liquid per unit cross-section area.

Similarly to other existing models \cite{Seker2008,Liu2016,Mercuri2017}, our approach postulates a sharp imbibition front that separates full pores from empty pores (Figure \ref{fig : ExpSetup}a, inset).
Our experiments rather indicate that a partially filled zone exists outside of the liquid-saturated zone.
For simplicity, we model the system using a sharp front at a position $L(t)$, which corresponds to the transition point between these two zones (saturated and partially filled), as defined experimentally (see \emph{Materials and Methods}).
The driving force $\DeltaPc$ then represents the capillary pressure at that transition point, which we assume constant. In fact, as discussed previously, the value of $\DeltaPc$ should be close to $\left( \DeltaPc \right)_\mathrm{eq}$ (see \emph{Materials and Methods}) because of the proximity to the partially filled zone.

With this sharp front assumption, we also ignore that a fraction of the liquid from the imbibition flow feeds the partially filled zone, which can be thought of as equivalent to an additional, effective evaporation loss from the halo.
Because evaporation slows down the propagation of the capillary front, the sharp front approximation should result in an overestimation of the speed of imbibition.
In other words, if this effect is significant, experimental data should display slower dynamics than predicted by our theory.

\subsection{Constants and definitions}

In cases where there is no evaporation, one-dimensional imbibition dynamics is driven by a uniform pressure gradient, $\nabla P = \DeltaPc / L(t)$, and follows the Lucas-Washburn (LW) equation \cite{Vincent2016,Vincent2017}
\begin{equation}
    L(t) = \sqrt{w t}
\label{eq : LW}
\end{equation}
with the LW coefficient
\begin{equation}
    w = \frac{2 \kappa | \DeltaPc |}{\phi}
\label{eq : LWcoeff}
\end{equation}
where $\phi$ is the porosity \cite{Vincent2016,Vincent2017}.
The coefficient $w$ has dimensions of a diffusivity (\unit{\meter\squared\per\second}).

In order to model evaporation from the wet porous zone, we assume like other authors \cite{Mercuri2017,Liu2016} that the mass flow rate due to evaporation from the halo is proportional to the halo area, and can be characterized by a evaporation rate per unit area, $q_\mathrm{evap}$ (\unit{\kg\per\m\squared\per\second}). A typical timescale can be constructed from this evaporation rate:
\begin{equation}
    \tau = \frac{\phi \rho H}{q_\mathrm{evap}},
\label{eq : tau}
\end{equation}
which corresponds to the hypothetical time it would take to completely empty an initially saturated zone of the porous medium under the sole effect of evaporation.

The evaporation flux $q_\mathrm{evap}$ depends on mass transfer in the air surrounding the sample, in particular on its relative humidity.
For now we only assume that it is constant in space and time, without making further assumptions on its value.
All results will be expressed as a function of $\tau$, which abstracts the evaporation rate (Equation \ref{eq : tau}).
Later, we will discuss how evaporation rates depends on physical parameters (relative humidity, boundary layer thickness etc., and on the Kelvin effect, see \emph{Halo evaporation rate} below).

Finally, we construct a typical lengthscale,
\begin{equation}
    L^\ast = \sqrt{w \tau}.
    \label{eq : TypicalLength}
\end{equation}
which illustrates the competition between imbibition ($w$) and evaporation ($\tau$)
As we will show below, $L^\ast$ identifies with the maximum spatial extent of the halo predicted by the one-dimensional approach, and also defines the halo dimensions in the two-dimensional model.

\subsection{1D halo dynamics}

When the extension of the halo is sufficiently small, $L \ll \idx{R}{drop}$, one can consider that the annular halo is equivalent to a rectangular zone of width, $W = 2 \pi \idx{R}{drop}$ and of length, $L(t)$. We define an equivalent one-dimensional axis $x$ in the direction of $L(t)$ with its origin at the edge of the droplet (see Figure \ref{fig : ExpSetup}b).

\begin{figure}[]
	\centering
	\includegraphics[scale=0.9]{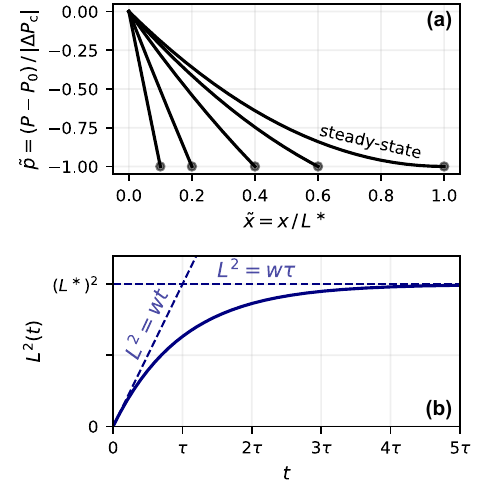}
    \caption{
		\small
		1D model for halo dynamics.
		(a) Pressure field (Equation \ref{eq : PressureFieldDimensionless}) for different position of the imbibition front ($\tilde{L}=0.1$, $0.2$, $0.4$, $0.6$ and $1$ from left to right where $\tilde{L} = L / L^\ast$ ; $\tilde{L}=1$ correspond to the steady-state while $\tilde{L} < 1$ represent a growing halo).
		(b) Corresponding solution for the position of the front squared as a function of time (Equation \ref{eq : FrontEquationSquared_1D}, continuous line); limiting regimes (early time and steady-state) are also displayed as dashed lines.
	}
	\label{fig : Theory_1D}
\end{figure}

In one dimension, Darcy's law writes
\begin{equation}
	\label{eq : Darcy}
	\idx{q}{p} = - \rho \kappa \pdv{P}{x}
\end{equation}
for the horizontal mass flux, $\idx{q}{p}(x)$, as a function of the pressure gradient.
Conservation of mass imposes that the difference in $\idx{q}{p}$ in the porous medium between positions $x$ and $x + \odif{x}$ should be equal to the mass loss by evaporation on the top surface of the medium on the corresponding infinitesimal surface: $H \times \left( \idx{q}{p}(x) - \idx{q}{p}(x + \odif{x})\right) = q_\mathrm{evap} \odif{x} $.
Combined with Darcy's law (Equation \ref{eq : Darcy}), we obtain
\begin{equation}
    \pdv[order=2]{P}{x} = \frac{q_\mathrm{evap}}{\rho \kappa H}.
\label{eq : PressureEquation}
\end{equation}
We define $\tilde{x} = x / L^\ast$ a reduced spatial coordinate, $\tilde{L} = L / L^\ast$ a reduced front position and $\tilde{p} = (P - P_0) / | \DeltaPc |$, a reduced pressure which equals 0 at the droplet ($\tilde{x}=0$), and $-1$ at the imbibition front ($\tilde{x}=1$).
With these dimensionless quantities, Equation \ref{eq : PressureEquation} rewrites
\begin{equation}
    \pdv[order=2]{\tilde{p}}{\tilde{x}} = 2.
\label{eq : PressureEquationDimensionless}
\end{equation}
Using the boundary conditions $\tilde{p}(0) = 0$ and $\tilde{p}(\tilde{L}(t)) = -1$, integration of this equation yields the pressure field
\begin{equation}
    \tilde{p}(\tilde{x}, t) = \tilde{x}^2 - \left( \tilde{L}(t) + \frac{1}{\tilde{L}(t)} \right) \tilde{x}.
\label{eq : PressureFieldDimensionless}
\end{equation}
We note that due to evaporation, the pressure distribution along the halo is not linear (i.e., uniform gradient) like in the standard Lucas-Washburn case, but parabolic (see Figure \ref{fig : Theory_1D}a).
This parabolic dependence was noted by Seker et al. \cite{Seker2008}; however these authors only considered the steady-state position of the halo. Below, we derive the full equation governing $L(t)$ with the evolving parabolic pressure field described by Equation \ref{eq : PressureFieldDimensionless}.

Differentiation of Equation \ref{eq : PressureFieldDimensionless} allows us to calculate the pressure gradient at the imbibition front
\begin{equation}
	\pdv{\tilde{p}}{\tilde{x}}_{\tilde{x} = \tilde{L}} = \tilde{L} - \frac{1}{\tilde{L}},
\label{eq : PressureFieldDerivativeDimensionless}
\end{equation}
which dictates how much the front advances, because it determines the liquid flux at the imbibition front from Darcy's law (Equation \ref{eq : Darcy}).
From Equation \ref{eq : PressureFieldDerivativeDimensionless}, this gradient vanishes when $\tilde{L} = 1$ (see also Figure \ref{fig : Theory_1D}a), in other words when $L = L^\ast$.
As a result, $L^\ast$ corresponds to the final steady-state position of the imbibition front, $\idx{L}{max}$, which results from the dynamic equilibrium between imbibition and evaporation:
\begin{equation}
	\label{eq : Lmax}
	\idx{L}{max} = L^\ast
\end{equation}

When $\tilde{L} < 1$, Equation \ref{eq : PressureFieldDerivativeDimensionless} predicts $\left( \pdv{\tilde{p}}/{\tilde{x}} \right)_{\tilde{x} = \tilde{L}} < 0$ (see also Figure \ref{fig : Theory_1D}a). From Darcy's law (Equation \ref{eq : Darcy}), this negative pressure gradient results in a positive, pressure-driven flow at the imbibition front, $\idx{q}{p}(L) > 0$, which itself makes the front grow at a velocity $\odv{L}/{t}$.
Conservation of mass $\idx{q}{p}(L) = \phi \rho \odv{L}/{t}$ combined with Equations \ref{eq : Darcy} and \ref{eq : PressureFieldDerivativeDimensionless} yields the differential equation
\begin{equation}
    \odv{\tilde{L}}{t} = \frac{1}{2 \tau} \left(\frac{1}{\tilde{L}} - \tilde{L} \right).
\label{eq : FrontDifferentialEquation_1D}
\end{equation}
which integrates readily with the initial condition $\tilde{L}(t=0) = 0$ to yield
$\tilde{L}^2(t) = 1 - \exp\left( -t / \tau \right)$
or, in dimensional form:
\begin{equation}
    L^2 (t) = w \tau \left( 1 - \exp \left( -\frac{t}{\tau} \right) \right).
	\label{eq : FrontEquationSquared_1D}
\end{equation}

Figure \ref{fig : Theory_1D}b presents a graph of Equation \ref{eq : FrontEquationSquared_1D} along with two limiting regimes.
At early times ($t \ll \tau$), $L(t) \simeq \sqrt{w t}$, as can be shown using a first-order Taylor expansion of Equation \ref{eq : FrontEquationSquared_1D} in $t / \tau$;
in other words, the classical LW equation (Equation \ref{eq : LW}) is recovered at short times.
This behavior is expected because initially the halo area is small, leading to a negligible evaporation flow rate from the pores, compared to capillary pumping.
The other limiting regime is the steady state (fixed front position) at large times, $L = L^\ast = \sqrt{w \tau}$, obtained for $t \gg \tau$, and which results from the dynamic equilibrium between capillary pumping and evaporation.
The two limiting regimes intersect at $t = \tau$ (see Figure \ref{fig : Theory_1D}b).

Note that an equation seemingly identical to Equation \ref{eq : FrontEquationSquared_1D} was found by Mercuri et al. \cite{Mercuri2017}.
Their approach consisted in comparing the evaporation flow rate to an hypothetical imbibition flow rate that would have the same characteristics as one without evaporation, e.g. having a linear pressure field (uniform pressure gradient) between the drop and the imbibition front.
As we showed above, mass conservation between imbibition and evaporation flows rather imposes a parabolic pressure profile.
Interestingly, the linear pressure profile hypothesis results in the same expression than Equation \ref{eq : FrontEquationSquared_1D}, but with different coefficients: $L^2(t) = L_\mathrm{m}^2 (1 - \exp(-t / \tau_\mathrm{m}))$, with $L_\mathrm{m} = L^\ast / \sqrt{2}$ and $\tau_\mathrm{m} = \tau / 2$.
In other words, Mercuri et al.'s approach underestimates both the final extent of the halo and the timescale to reach steady-state by a factor $\sqrt{2}$ and $2$, respectively.

\subsection{2D halo dynamics}

When the halo dimensions become non negligible compared to the size of the droplet, the 1D calculations developed above are expected to be less accurate.
Below we evaluate the corrections needed to the one-dimensional model and derive a full set of equations describing the dynamics for any halo dimension, assuming that the imbibition front has a circular shape that grows radially, with a radius $R(t) = \idx{R}{drop} + L(t)$
measured from the center of the drop (see Figure \ref{fig : ExpSetup}).

Liu et al. \cite{Liu2016} established a differential equation describing radial imbibition dynamics in the presence of evaporation. Using our definitions of $w$ and $\tau$ (Equations \ref{eq : LWcoeff} and \ref{eq : tau}, respectively) their governing equation for the radius of the halo, $R(t)$, rewrites
\begin{equation}
	\label{eq : FrontDifferentialEquation_2D}
	\scriptsize
	2 R \odv{R}{t} \ln\left(\frac{R}{\idx{R}{drop}} \right)
	+ \frac{1}{2 \tau} \left( \idx{R}{drop}^2 - R^2 + 2 R^2 \ln \left(  \frac{R}{\idx{R}{drop}} \right) \right)
	 = w.
\end{equation}
Liu et al. \cite{Liu2016} did not provide an analytical solution to Equation \ref{eq : FrontDifferentialEquation_2D} and used numerical methods to solve for $R(t)$.
Below, we show that semi-analytical solutions can be found.
First, we will discuss the general shape of solutions to Equation \ref{eq : FrontDifferentialEquation_2D} and limiting cases, which will help us establishing analytical solutions for the dynamics at all times.

\begin{figure}[]
	\centering
	\includegraphics[scale=0.9]{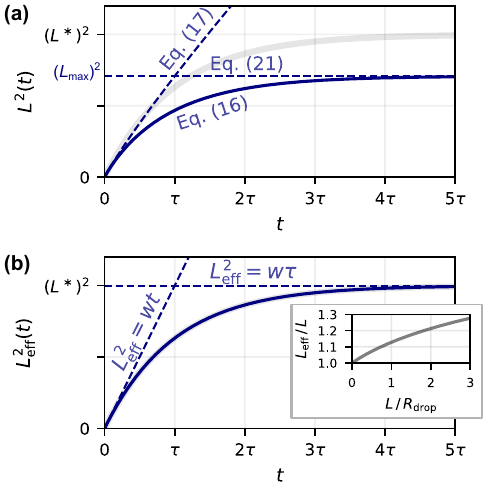}
    \caption{
		\small
		2D model for halo dynamics.
		(a) Numerical solution for the square of the position of the front as a function of time $L^2(t) = (R(t) - \idx{R}{drop})^2$ (Equation \ref{eq : FrontDifferentialEquation_2D}, continuous blue line, solved in the case $L^\ast / \idx{R}{drop} = 2$).
		Analytical solutions for limiting regimes (early time, Equation \ref{eq : FrontDynamics2D_Init} and steady-state, Equation \ref{eq : Rmax}) are also displayed as dashed lines.
		The continuous pale gray line is the prediction from the 1D model (Equation \ref{eq : FrontEquationSquared_1D}) for comparison.
		(b) Data in (a) plotted in terms of the effective halo size, $\idx{L}{eff}$ (Equations \ref{eq : Leff}-\ref{eq : Leff_L}):
		continuous blue line: numerical simulation, continuous light gray line (indistinguishable from the previous one): analytical solution (Equation \ref{eq : FrontEquationSquared_2D_Leff}), dashed lines: limiting cases (Equations \ref{eq : FrontDynamics2D_Init_Leff} and \ref{eq : Leff_max}).
		This graph is identical to Figure \ref{fig : Theory_1D}b, replacing $L$ by $\idx{L}{eff}$.
		Inset: ratio between effective halo extension $\idx{L}{eff}$ and actual halo extension $L$, as a function of relative size between halo extension and droplet radius, calculated using Equation \ref{eq : Leff_L}.
	}
	\label{fig : Theory_2D}
\end{figure}

An example of a numerical solution to Equation \ref{eq : FrontDifferentialEquation_2D} is shown in Figure \ref{fig : Theory_2D}a (continuous blue line), for the case where $L^\ast = \sqrt{w \tau} = 2 \times \idx{R}{drop}$, i.e., where the final halo width ($\idx{L}{max} \simeq L^\ast$ as predicted from the one-dimensional model) should be approximately twice as large as the droplet radius; this case is close to the largest halo sizes observed in our experiments (see \emph{Results and Discussion}).
Figure \ref{fig : Theory_2D}a shows significant deviations from the 1D solution in this situation (Figure \ref{fig : Theory_2D}a, pale gray line).
In particular, the final extension of the front is noticeably smaller than predicted by the one-dimensional approach  ($\idx{L}{max} < L^\ast$).

When $t \ll \tau$, the second term in Equation \ref{eq : FrontDifferentialEquation_2D} is negligible, and the equation integrates into
\begin{equation}
	\label{eq : FrontDynamics2D_Init}
	R^2 \ln \left( \frac{R}{\idx{R}{drop}} \right) + \frac{\idx{R}{drop}^2 - R^2}{2} = wt,
\end{equation}
which corresponds to the dynamics of 2D radial imbibition without evaporation as already established in the literature \cite{Hyvaluoma2006}.
The prediction of Equation \ref{eq : FrontDynamics2D_Init} is shown as a dashed line in Figure \ref{fig : Theory_2D}a, showing that it indeed reproduces well the early times dynamics.
\footnote{
	Also note that equation \ref{eq : FrontDynamics2D_Init} further reduces to the LW equation when $L(t) = R - \idx{R}{drop} \ll \idx{R}{drop}$, i.e., $L^2(t) \simeq w \times t$, as can be verified with a first-order Taylor expansion in $L / \idx{R}{drop}$.
}
Equation \ref{eq : FrontDynamics2D_Init} can be rewritten
\begin{equation}
	\label{eq : FrontDynamics2D_Init_Leff}
	\idx{L}{eff}(t) = \sqrt{w  t}
\end{equation}
where we have defined an effective halo length $\idx{L}{eff}$
\begin{equation}
	\label{eq : Leff}
	\idx{L}{eff} = \sqrt{ R^2 \ln \left( \frac{R}{\idx{R}{drop}} \right) + \frac{\idx{R}{drop}^2 - R^2}{2}}
\end{equation}
which incorporates 2D effects in the halo dynamics, and can be calculated knowing $\idx{R}{drop}$ and $R(t) = \idx{R}{drop} + L(t)$.
Equivalently,
\begin{equation}
	\label{eq : Leff_L}
	\idx{L}{eff} = \idx{R}{drop} \sqrt{ \left( 1 + \frac{L}{\idx{R}{drop}} \right)^2 \left[ \ln \left( 1 + \frac{L}{\idx{R}{drop}} \right) - \frac{1}{2} \right] + \frac{1}{2}}.
\end{equation}

The second limiting case is the steady-state radius of the halo, $\idx{R}{max} = \idx{R}{drop} + \idx{L}{max}$, which is obtained by setting $\odv{R}/{t} = 0$ in Equation \ref{eq : FrontDifferentialEquation_2D}, resulting in
\begin{equation}
	\label{eq : Rmax}
	\idx{R}{drop}^2 - \idx{R}{max}^2 + 2 \idx{R}{max}^2 \ln \left( \frac{\idx{R}{max}}{\idx{R}{drop}} \right) = 2 w \tau.
\end{equation}
Equation \ref{eq : Rmax} reduces to Equation \ref{eq : Lmax}, i.e., $\idx{L}{max} = \sqrt{w \tau}$ when $\idx{L}{max} = \idx{R}{max} - \idx{R}{drop} \ll \idx{R}{drop}$, as can be shown using a second-order Taylor expansion in $\idx{L}{max} / \idx{R}{drop}$ (the first order cancels out).
The steady-state solution predicted by Equation \ref{eq : Rmax} is plotted as a horizontal dashed line in Figure \ref{fig : Theory_2D}a and matches the numerical simulation at large times.
Interestingly, $\idx{L}{eff}$ appears again here, because Equation \ref{eq : Rmax} can be rewritten
\begin{equation}
	\label{eq : Leff_max}
	L_\mathrm{eff}^{(\mathrm{max})} = L^\ast
\end{equation}
where $L_\mathrm{eff}^{(\mathrm{max})} = \idx{L}{eff}(L = \idx{L}{max})$ as calculated from Equation \ref{eq : Leff_L}, and where we recall the definition $L^\ast = \sqrt{w \tau}$ (Equation \ref{eq : TypicalLength}).

Since the quantity $\idx{L}{eff}(t)$ defined in Equations \ref{eq : Leff} or \ref{eq : Leff_L} has the same limiting regimes (Equations \ref{eq : FrontDynamics2D_Init_Leff} and \ref{eq : Leff_max}) than $L(t)$ in the one-dimensional approach (Equations \ref{eq : LW} and \ref{eq : Lmax}), one may suspect that $\idx{L}{eff}$ in the 2D case follows the same equation as $L$ in the 1D case, for all times.
This is actually the case, as can be demonstrated by injecting Equation \ref{eq : Leff} in the differential equation (Eq. \ref{eq : FrontDifferentialEquation_1D}), which yields
\begin{equation}
	\label{eq : FrontDifferentialEquation_2D_Leff}
	\odv{\tilde{L}_\mathrm{eff}}{t} = \frac{1}{2 \tau} \left( \frac{1}{\tilde{L}_\mathrm{eff}} - \tilde{L}_\mathrm{eff}  \right)
\end{equation}
where $\tilde{L}_\mathrm{eff} = \idx{L}{eff} / L^\ast$.
Since Equation \ref{eq : FrontDifferentialEquation_2D_Leff} is the same as Equation \ref{eq : FrontDifferentialEquation_1D}, the solution is identical:
\begin{equation}
	\label{eq : FrontEquationSquared_2D_Leff}
	\idx{L}{eff}^2(t) = w \tau \left( 1 - \exp \left( - \frac{t}{\tau} \right) \right)
\end{equation}
which indeed describes perfectly the numerical solutions to Equation \ref{eq : FrontDifferentialEquation_2D}, see Figure \ref{fig : Theory_2D}b (gray curve, indistinguishable from the numerical simulation in blue).
As a result, the 2D dynamics follows the same equations as the 1D model, provided one uses an effective imbibition length $\idx{L}{eff}$ instead of the geometrical imbibition length, $L$; $\idx{L}{eff}$ and $L$ are close when $L$ is small compared to the droplet radius, but get increasingly different as $L / \idx{R}{drop}$ increases (Figure \ref{fig : Theory_2D}b, inset).

In practice, if $R(t) = L(t) - \idx{R}{drop}$ is known, e.g. from experimental measurements, comparison with Equation \ref{eq : FrontEquationSquared_2D_Leff} can be done directly by calculating $\idx{L}{eff}$ from $R(t)$ using Equation \ref{eq : Leff} or Equation \ref{eq : Leff_L}.
If on the contrary one wants to calculate $L(t)$ from the analytical expression of $\idx{L}{eff} (t)$, Equations \ref{eq : Leff} or \ref{eq : Leff_L} must be inverted, which requires e.g. numerical approaches because they are implicit for $R(t)$ or $L(t)$.
Such numerical solving is however much more direct than having to perform a complete numerical simulation of the initial differential equation (Equation \ref{eq : FrontDifferentialEquation_2D}).


\subsection{Halo evaporation rate}

The general solutions to our theory of imbibition halo dynamics developed above are expressed as a function of the grouped parameter $\tau$, which directly depends on the evaporation rate by unit area (flux) in the halo, $q_\mathrm{evap}$ (Equation \ref{eq : tau}).
Here we discuss physical expectations for $q_\mathrm{evap}$, and its dependence with respect to the humidity of the air surrounding the sample.

It is common to model evaporation from porous media using the concept of an effective boundary layer of thickness, $\delta$ (\unit{\m}) above the surface where evaporation occurs, where gradients of water vapor concentration in air are localized \cite{Nobel2020}.
Within this framework, transport is limited by vertical diffusion of water vapor within that boundary layer, assumed to be uniform in thickness.
As a result, from Fick's law, the evaporation flux (mass flow rate of evaporation per unit area) is:
\begin{equation}
    q_\mathrm{evap}^\mathrm{(halo)} = \epsilon \rho D \frac{\Delta \rh}{\delta}
\label{eq : BoundaryLayerFlux}
\end{equation}
where $D$ is the diffusivity of water vapor in air (\unit{\meter\squared\per\second}), and
\begin{equation}
	\label{eq : DeltaH_Halo}
	\Delta \rh = \idx{\rh}{eq} - \rh
\end{equation}
is the difference of relative humidity between the equilibrium RH of the liquid at the evaporative surface (liquid-vapor menisci in the pores), $\idx{\rh}{eq}$, and the humidity $\rh$ imposed in the air far away from the surface.
Due to the Kelvin effect, $\idx{\rh}{eq} < 1$ is lowered compared to bulk saturation ($\idx{\rh}{sat} = 1$), see \emph{Materials and Methods}.
The dimensionless parameter
\begin{equation}
	\label{eq : epsilon}
	\epsilon = \vm \psat / (\idealR T)
\end{equation}
depends on the physical properties of water and is on the order of \num{2e-5} at room temperature (see \emph{Supporting Information}).

The effective boundary layer thickness, $\delta$, depends on the dimensions of the evaporative surface and on the properties of the air flow above the sample.
Boundary layer theory indicates that for an object of typical length, $\ell$ (\unit{\m}) in an air flow of velocity, $v$ (\unit{\meter\per\second}),
\begin{equation}
	\label{eq : BoundaryLayerThickness}
	\delta \simeq \alpha \sqrt{\frac{\ell}{v}}
\end{equation}
with $\alpha \sim \numrange{4e-3}{6e-3}$ \unit{\m} \unit{\second\tothe{\text{-\sfrac12}}} at ambient temperatures \cite{Nobel2020}.

Note that Equation \ref{eq : BoundaryLayerFlux} reduces a complex 3-dimensional problem into an equivalent one-dimensional formulation.
As such, the quantities $q_\mathrm{evap}$ and $\delta$ need to be considered as spatially averaged.
More generally, when analyzing our results using Equations \ref{eq : BoundaryLayerFlux}-\ref{eq : DeltaH_Halo}, the extracted evaporation rates and equilibrium humidities should be interpreted as averaged over the halo surface, e.g. because the Kelvin effect may vary spatially due to the pressure gradient between the droplet and the menisci at the imbibition front (see e.g Figure \ref{fig : Theory_1D}a and Equation \ref{eq : KelvinPc} relating liquid pressure to equilibrium vapor pressure).

\subsection{Droplet evaporation rate}

In order to account for volumetric variations of the droplet itself, we assume that evaporation dynamics is similar on a porous medium and on a flat, non-porous surface.
This hypothesis will be justified by our experimental results (see \emph{Results and Discussion}).
Sessile droplet evaporation is intrinsically a 3-dimensional problem, with highly nonuniform evaporation rates across the droplet surface \cite{Brutin2018}.
However, when the contact angle of the droplet on the surface remains small ($\theta < \qty{40}{\degree}$), evaporation rates of water have been shown to be well described by a simple formula \cite{Hu2002}:
\begin{equation}
    Q_\mathrm{evap}^\mathrm{(drop)} = 4 \idx{R}{drop} \epsilon \rho D \Delta \rh
\label{eq : DropletFlux}
\end{equation}
where $Q_\mathrm{evap}^\mathrm{(drop)}$ is the total mass evaporation rate ($\unit{kg/s}$).
Contrary to the previous case of evaporation from the pores, the liquid in the droplet is not confined and we expect its equilibrium humidity to be 100\%RH ($\idx{\rh}{sat} = 1$), i.e.,
\begin{equation}
	\label{eq : DeltaH_Drop}
	\Delta \rh = \idx{\rh}{sat} - \rh
\end{equation}

By comparing Equations \ref{eq : BoundaryLayerFlux} and \ref{eq : DropletFlux}, and defining an effective drop evaporation flux $q_\mathrm{drop}$ such that
\begin{equation}
	\label{eq : qdrop}
	\idxy{q}{evap}{drop} = \frac{Q_\mathrm{evap}^\mathrm{(drop)}}{\pi \idx{R}{drop}^2},
\end{equation}
one finds that evaporation from a sessile droplet can also be described by an equation similar to Equation \ref{eq : BoundaryLayerFlux}, i.e., $\idxy{q}{evap}{drop} = \epsilon \rho D \Delta \rh / \delta_\mathrm{drop}$, with
\begin{equation}
    \delta_\mathrm{drop} = \frac{\pi}{4} \idx{R}{drop},
\label{eq : EffectiveDeltaDrop}
\end{equation}
an equivalent boundary layer thickness, which represents the value of $\delta$ that would produce the same evaporation rate as a 3D droplet in a one-dimensional situation.
Equation \ref{eq : EffectiveDeltaDrop} illustrates that 3D vapor diffusion around sessile droplets occur on typical lengthscales set by the droplet base radius.

\section{Results and Discussion}

Below, we describe our experimental results in the light of the theoretical developments described above.
First, we provide a qualitative description of the different stages of droplet dynamics from deposition on the porous surface until complete disappearance by evaporation.
Then, we focus on the dynamics of expansion of the wetted annulus around the droplet (the halo), and on its dependence on the imposed RH.
We also characterize the evaporation dynamics of the droplet itself, and discuss implications of all results in terms of evaporation rates and competing mechanisms of transport.

\subsection{Complete dynamics}

\begin{figure}
	\centering
	\includegraphics[scale=0.105]{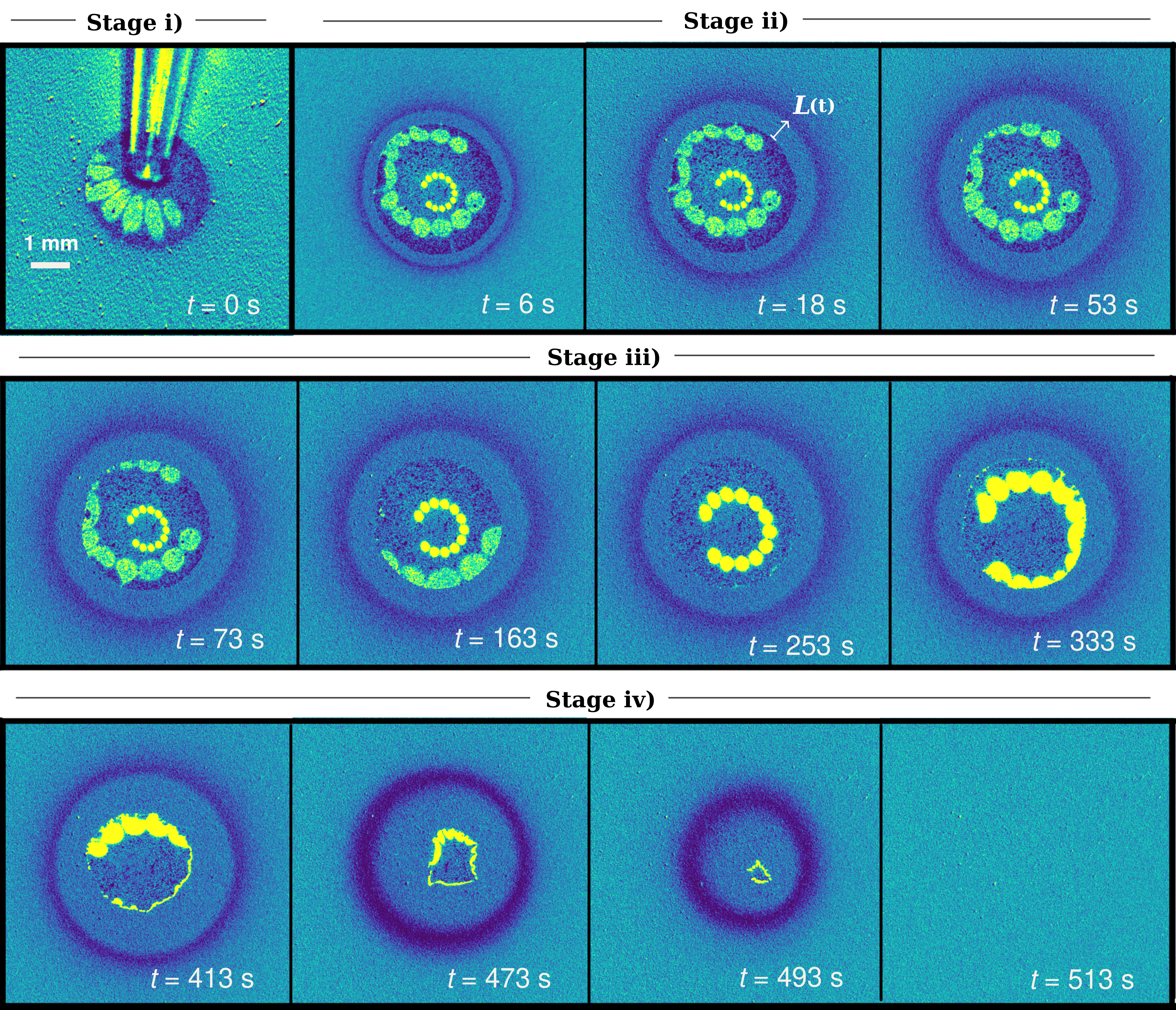}
    \caption{
		\small
		Complete dynamics after deposition on the mesoporous substrate of a \qty{1}{\micro\liter} water droplet, at $\rh = \qty{30}{\percent}$RH.
		After droplet deposition with the micro-pipette (first image), we observe 4 stages in the dynamics:
		i) fast spreading,
		ii) halo growth,
		iii) steady-state halo and droplet evaporation,
		iv) recession of droplet and halo.
		Note that time intervals between images are not constant due to various time scales in the dynamics.
		Images are background-subtracted and displayed in pseudocolors as in Figure \ref{fig : ExpSetup}b.
	}
    \label{fig : ImageSequence}
\end{figure}

After droplet deposition on the porous substrate, we observed four successive stages: i) droplet spreading, ii) halo development, iii) steady-state halo combined with droplet flattening, iv) simultaneous droplet and halo recession until complete evaporation of the liquid (Figure \ref{fig : ImageSequence}).

In \emph{stage i)}, we observed fast spreading after droplet deposition, until relaxation to a state with a stable contact angle of \qty{27(2)}{\degree} (see \emph{Materials and Methods}).
Spreading is known to occur at time scales of $\sim$ms \cite{Eddi2013,Davis1999}, far below the temporal resolution of our images, and happens between the first two images in Figure \ref{fig : ImageSequence}.
Note that we estimate that \emph{vertical} imbibition into the porous medium below the droplet its contact surface should occur at similar time scales (see \emph{Supporting Information}).
After stage i), the droplet became pinned to the substrate and evolved with a fixed contact line during stages ii) and iii).

\emph{Stage ii)} consisted in the spontaneous, horizontal imbibition of the liquid from the droplet into the pores by capillary action, forming an expanding halo of wetted material in the porous medium around the drop.

In \emph{stage iii)} the halo reached a steady-state position because of a dynamic equilibrium between capillary flow and evaporation.
At the same time, the droplet also continuously lost volume by evaporation at its top surface.
Since the droplet contact line was pinned, droplet evaporation resulted in a progressive decrease of the contact angle.
This flattening of the droplet is visible in the images (second line in Figure \ref{fig : ImageSequence}) as expanding ring light reflections (see \emph{Materials and Methods} and \emph{Supporting Information}).

Finally in \emph{stage iv)}, the contact line detached and the drop retracted.
This stage corresponds to a regime of evaporation with constant contact angle and receding contact line \cite{Brutin2018} until complete droplet evaporation.
During droplet retraction, the halo position followed the contact line, but not necessarily with the same speed, as evidenced with the apparent broadening of the halo size before complete disappearance (Figure \ref{fig : ImageSequence}, last line).
This delayed halo retraction could be related to the hysteresis between filling and evaporation (Figure \ref{fig : Isotherm}), which might also explain the visible darkening of the partially filled zone around the halo during droplet retraction.
We have not studied this dynamics in detail, and will not discuss it further in the present paper.

In the following, we focus our analysis on the halo and droplet dynamics in regimes ii) and iii), corresponding to the situation described in the \emph{Theory} section.

\subsection{Halo expansion dynamics}

\begin{figure}[]
	\centering
	\includegraphics[scale=0.9]{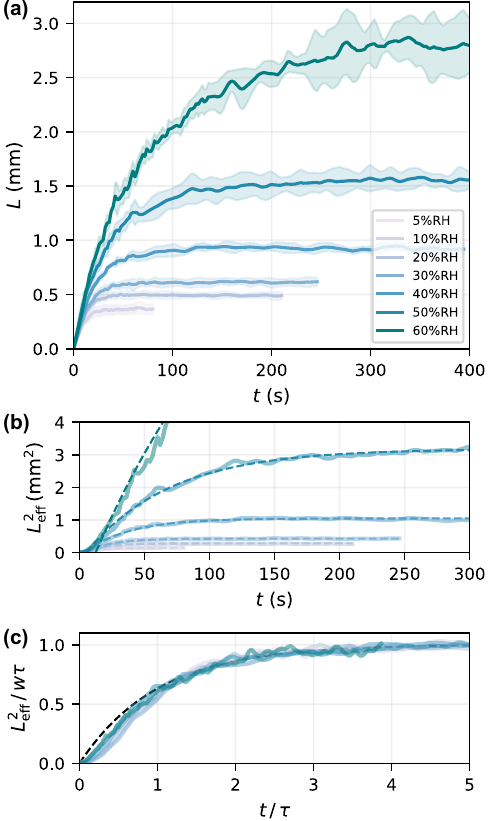}
    \caption{
		\small
		Dynamics of the imbibition front (halo).
		(a) Measured extension, $L(t)$, of the halo as a function of time; $t=0$ corresponds to droplet deposition time.
		Colors represent experiments at different relative humidities: continuous lines display average behavior while the shaded areas show the observed variations (standard deviation) across $5 \pm 1$ repeats.
		Note that the curves for \qty{5}{\percent}RH and \qty{10}{\percent}RH overlap.
		(b) Fitting of the experimental halo dynamics from (a) with Equation \ref{eq : FrontEquationSquared_2D_Leff}, yielding the parameters $w$ and $\tau$.
		The effective imbibition length, $\idx{L}{eff}(t)$ is calculated from $L(t)$ (Equation \ref{eq : Leff_L}).
		For clarity, only the average lines from Figure \ref{fig : Halo_Dynamics}a are shown (continuous lines);
		dashed lines are the fitted curves.
		(c) Rescaled experimental data using $\tau$ and $w$ obtained from the fits in (b).
		The black, dashed line corresponds to Equation \ref{eq : FrontEquationSquared_2D_Leff}.
	}
	\label{fig : Halo_Dynamics}
\end{figure}

We systematically studied how the dynamics of halo formation (Figure \ref{fig : ImageSequence}, stages ii) and iii)) depended on the RH of the air surrounding the droplet and mesoporous samples.
Figure \ref{fig : Halo_Dynamics}a presents the temporal evolution of the halo size (imbibition front position, $L(t)$) for a variety of humidities between 5\%RH and 60\%RH.
Time $t=0$ corresponds to deposition of the droplet on the nanoporous sample;
since we could not resolve the initial phase of droplet spreading, $t=0$ also corresponds within our measurement accuracy to the start of the imbibition process from the pinned contact line of the droplet.
At each RH, we repeated the experiment 4 to 6 times, and the results in Figure \ref{fig : Halo_Dynamics}a indicate the average behavior and typical variation (standard deviation) across repeats.

At large RH ($>60\--70$\%RH), the halo was not visible or hard to distinguish, because the porous surface should already be completely or partially filled with liquid water prior to deposition of the droplet, due to spontaneous capillary condensation of water vapor into the pores (see \emph{Materials and Methods} and Figure \ref{fig : Isotherm}).

\begin{figure}[]
	\centering
	\includegraphics[scale=0.9]{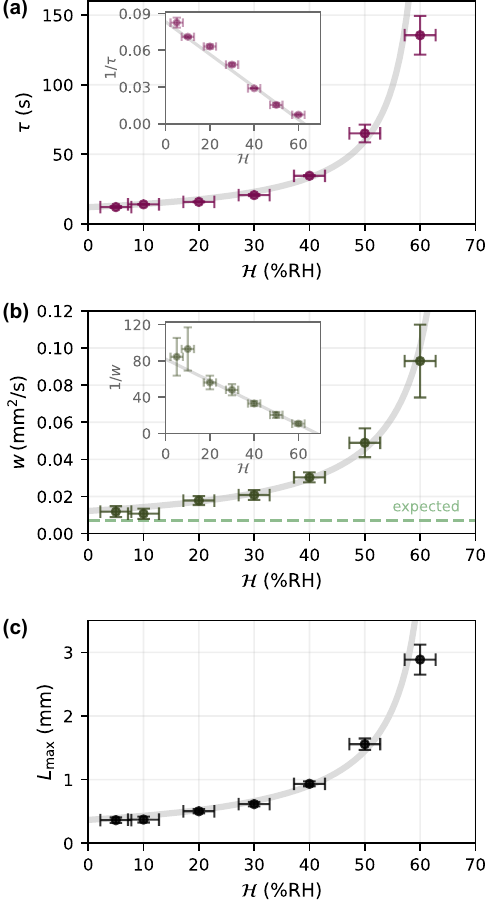}
    \caption{
		\small
		Halo characteristics as a function of the imposed RH.
		Values are extracted from the fitting of the data in Figure \ref{fig : Halo_Dynamics} using Equation \ref{eq : FrontEquationSquared_2D_Leff}.
		(a) Characteristic evaporation time, $\tau$.
		Inset: graph of $1 / \tau$.
		(b) Lateral transport coefficient (LW parameter, $w$).
		The green line shows the expected value of $w$ based on material properties of the porous medium (Equation \ref{eq : LW}).
		Inset : graph of $1 / w$.
		(c) Maximal (steady-state) extension, $\idx{L}{max}$.
		In all panels, gray lines are fits to the data, using Equation \ref{eq : FitFormula_Tau} for $\tau$ (panel a), Equation \ref{eq : FitFormula_w} for $w$ (panel b), and both previous equations for $\idx{L}{max}$ (panel c).
		See text for details.
	}
	\label{fig : Halo_FitResults}
\end{figure}

At 60\%RH and below, we could clearly see a halo develop, with an initial expansion phase followed by a steady-state (see Figure \ref{fig : Halo_Dynamics}a).
The maximum, steady-state extension of the front, $\idx{L}{max}$, strongly depended on the imposed RH, and increased from approximately $0.3$ mm at low humidities to $3$ mm at the maximum investigated RH of 60\%RH (see Figure \ref{fig : Halo_Dynamics}a and Figure \ref{fig : Halo_FitResults} further below).

Qualitatively, the dynamics shown in Figure \ref{fig : Halo_Dynamics}a corresponds to that expected from theory, with a halo that is first expanding, then stopping due to a dynamic equilibrium between capillary pumping (from droplet to halo) and evaporation (from halo to air).
At higher RH, the dynamic equilibrium is modified due to reduced evaporation: capillary suction "wins" over evaporation, and the front is able to reach larger dimensions.

Quantitatively, we expect this imbibition/evaporation competition to result in an exponential growth towards a steady-state.
For small halo dimensions, the quantity following this exponential growth is $L^2(t)$ (see Equation \ref{eq : FrontEquationSquared_1D}), with a time constant, $\tau$, depending on the evaporation rate (Equation \ref{eq : tau}).
For larger dimensions, when $L(t)$ is no longer negligible compared to the droplet base radius $\idx{R}{drop}$, $L(t)$ needs to be replaced by $\idx{L}{eff}(t)$, which follows an identical equation (Equation \ref{eq : FrontEquationSquared_2D_Leff}).
The quantity $\idx{L}{eff}$ is an effective imbibition length (Equation \ref{eq : Leff} or \ref{eq : Leff_L}), which accounts for the two-dimensional geometry.
Since in our experiments, the halo expanded to dimensions of more than twice $R_\mathrm{drop}$, we use this second, two-dimensional approach to analyze our data.

Figure \ref{fig : Halo_Dynamics}b presents the evolution of $\idx{L}{eff}^2(t)$, calculated with Equation \ref{eq : Leff_L} from $L(t)$ in Figure \ref{fig : Halo_Dynamics}a (continuous lines).
We fitted this data with Equation \ref{eq : FrontEquationSquared_2D_Leff} using a least-squares method (dashed lines).
In order to best represent the data, we added a time origin, $t_0$ as a free parameter in the fitting procedure, which we found was always small compared to $\tau$ (see \emph{Supporting Information}).

From the fitting procedure, we extracted two quantities: the evaporation-related, transient time $\tau$ and the lateral imbibition coefficient (LW constant), $w$.
The fit quality was very good for all investigated RH, as illustrated by a rescaling of the data in time\footnote{
	We did not include $t_0$ in the rescaled data in Figure \ref{fig : Halo_Dynamics}c, i.e., the reduced time is $t / \tau$ and not $(t - t_0) / \tau$.
}
(normalized with $\tau$) and space ($\idx{L}{eff}^2$ divided by $w \tau$), following Equation \ref{eq : FrontEquationSquared_2D_Leff} (see Figure \ref{fig : Halo_Dynamics}c).
Small deviations are visible at early times, which could be due to mechanisms limiting radial transport for small halo dimensions that we have not considered in the model.
Since these deviations have little impact on our analysis, we do not consider them further.

Below, we analyze the dependence of the extracted halo parameters (transient evaporation timescale, $\tau$, radial LW transport coefficient, $w$) on the air RH.
We also discuss implications for the steady-state extension of the halo, $\idx{L}{max}$.

Figure \ref{fig : Halo_FitResults}a shows a strong increase of $\tau$ with rising RH.
This behavior is expected, because $\tau$ is inversely proportional to the evaporation rate by unit area, $q_\mathrm{evap}$ (Equation \ref{eq : tau}), which itself decreases as the air humidity increases (Equation \ref{eq : BoundaryLayerFlux}).
This dependence is better visible when plotting $1 / \tau$ (Figure \ref{fig : Halo_FitResults}a, inset), showing a linear decrease of $q_\mathrm{evap} \sim 1 / \tau$ with RH, as expected from Equation \ref{eq : BoundaryLayerFlux}.
Interestingly, $1 / \tau$ reaches zero (i.e., no evaporation from the halo) at a value of relative humidity, $\idx{\rh}{eq} \simeq \qty{63}{\percent}$RH, far below bulk saturation ($\rh_\mathrm{sat} = \qty{100}{\percent}$RH).
We attribute this observation to the Kelvin effect, i.e., a decrease of the equilibrium humidity of the liquid confined in the pores due to the curvature of the liquid-vapor interface (see \emph{Materials and Methods}).

More quantitatively, we fitted the data with the formula
\begin{equation}
	\label{eq : FitFormula_Tau}
	\tau = \frac{\tau_\mathrm{e}}{\idx{\rh}{eq} - \rh}
\end{equation}
where we have defined a characteristic time,
\begin{equation}
	\label{eq : FitFormula_Tau0}
	\tau_\mathrm{e} = \frac{\phi H \delta}{\epsilon D}
\end{equation}
as predicted from Equations \ref{eq : tau} and \ref{eq : BoundaryLayerFlux}.
From the fitted curves shown as gray lines in Figure \ref{fig : Halo_FitResults}a, we have extracted $\idx{\rh}{eq} = \qty{62.6(0.5)}{\percent}$ RH, and $\tau_\mathrm{e} = \qty{7.52(0.12)}{\second}$.

The value of $\idx{\rh}{eq}$ is within the range of the Kelvin equilibrium RH estimated from water sorption isotherms, $\idx{\rh}{eq} = \qty{59.0(4.3)} {\percent}$ RH and corresponding to radii of curvature of $\simeq \qty{2}{\nm}$ in the pores (see \emph{Materials and Methods}).
This good agreement indicates that effects such as variable curvature of the menisci across the halo does not play a large role in the average equilibrium RH across the whole halo surface (see \emph{Supporting Information} for further discussion).

Additionally, from the extracted value of $\tau_\mathrm{e}$, known material properties ($H$, $\phi$, see \emph{Materials and Methods}) and water properties ($\epsilon$, $D$, see \emph{Supporting Information}), we used Equation \ref{eq : FitFormula_Tau0} to calculate the effective boundary layer thickness that governs evaporation fluxes from the halo, $\delta = \qty{2.2(0.2)}{\milli\meter}$.
This value matches estimates from boundary layer theory (Equation \ref{eq : BoundaryLayerThickness}), $\delta \simeq \qtyrange{2}{3}{\mm}$, using the typical air flow velocity and sample dimensions (see \emph{Materials and Methods}).
Halo transient times and corresponding evaporation rates are thus well-described by our theory, based on boundary-layer diffusion and the Kelvin equation.


The lateral capillary transport coefficient, $w$ (see Figure \ref{fig : Halo_FitResults}b) also shows a strong dependence on RH, which this time is surprising: $w$ is a parameter that should depend only on properties of the porous medium (permeability, capillary pressure, etc. see Equation \ref{eq : LWcoeff}) and not on external conditions; it should thus be constant across all experiments.
Here, we observe on the contrary that the extracted $w$ increases by a factor $\simeq 8$ between \qty{5}{\percent}RH and \qty{60}{\percent}RH.

Since a graph of $1 / w$ as a function of RH indicates an approximate linear decrease with RH (see Figure \ref{fig : Halo_FitResults}b, inset), we fitted the data with the empirical equation
\begin{equation}
	\label{eq : FitFormula_w}
	w = w_0 \left(1 - \frac{\rh}{\rh_\mathrm{c}} \right)^{-1}
\end{equation}
where $w_0 = \qty{12(1)e-9}{\meter\squared\per\second}$ is the extrapolated LW coefficient at a relative humidity of 0\%RH, and $\rh_\mathrm{c} = \qty{68(3)}{\percent}$ RH is a critical relative humidity at which the LW coefficient seemingly diverges;
$\rh_\mathrm{c}$ is again similar to the confined equilibrium RH $\idx{\rh}{eq} = \qty{59.0(4.3)} {\percent}$ RH but closer to the RH of capillary condensation $\idx{\rh}{cond} = \qty{68.4(5.9)}{\percent}$ RH, as estimated from the water sorption isotherm (see \emph{Materials and Methods}).

Interestingly, $w_0$ is of the same order of magnitude, but larger than the expected value of $w$ based on material properties from Equation \ref{eq : LWcoeff}, $w = \qty{7(1)e-9}{\meter\squared\per\second}$ (see \emph{Supporting Information}), which is displayed as a horizontal dashed line in Figure \ref{fig : Halo_FitResults}b.
Thus the system dynamics apparently approaches that expected from capillary flow theory only at extremely low RH.

Potentially, our estimate of $w$ could be biased because of our experimental definition of the halo position, which ignores the partially filled zone that surrounds it (see \emph{Materials and Methods}).
However, such an approach should underestimate $w$ rather than overestimate it, because defining a front position in that external zone would lead to a larger front width.
Additionally, because the sharp-front approximation of our model excludes mass exchange between the halo and the partially filled zone, experimental values of $w$ should be lower than theoretically expected, not larger (see \emph{General considerations} in \emph{Theory}).

Another mechanism that could explain a large and varying $w(\rh)$ is partial adsorption/condensation of vapor into the pores prior to droplet deposition.
Indeed, sorption isotherms indicate RH-dependent mass uptake (Figure \ref{fig : Isotherm}), which may impact capillary flow by reducing the effective porosity of the sample and/or by modifying the meniscus shape and associated capillary pressure.
In section \emph{Effect of pore pre-filling}) below, we show that this effect is too weak to explain the observed variations of $w$.
The most likely explanation involves vapor transport in air along the surface of the sample, in parallel with the capillary flow (see \emph{Lateral vapor transport} below).


Finally, the parameters $w$ and $\tau$ studied previously directly impose the physical dimensions of the halo.
Indeed, the steady-state value of the effective imbibition front position $\idx{L}{eff}(t)$ is $\idx{L}{eff}^{(\mathrm{max})} = \sqrt{w \tau}$ (Equation \ref{eq : FrontEquationSquared_2D_Leff}, Figure \ref{fig : Halo_Dynamics}c).
The actual steady-state halo width $\idx{L}{max}$, is then obtained from $\idx{L}{eff}^{(\mathrm{max})}$ by numerically inverting Equation \ref{eq : Leff_L}.
Following this procedure, we extracted the experimental $\idx{L}{max}$ from the values of $\tau$ and $w$ in Figures \ref{fig : Halo_FitResults}a and \ref{fig : Halo_FitResults}b, respectively, see Figure \ref{fig : Halo_FitResults}c.
With the same method, we also combined the corresponding fit formulas (Equation \ref{eq : FitFormula_Tau}) for $\tau(\rh)$ and Equation \ref{eq : FitFormula_w} for $w(\rh)$ to generate the gray fitting curve in Figure \ref{fig : Halo_FitResults}c for $\idx{L}{max}(\rh)$.
As with the other quantities, $\idx{L}{max}$ shows a strong increase as a function of RH (Figure \ref{fig : Halo_FitResults}a), corresponding to a combination of the (expected) lowering of the evaporation rate with increasing RH, and of the (unexpected) simultaneous increase of the lateral transport coefficient.

\subsection{Droplet evaporation dynamics}

We also investigated the evaporation dynamics of the droplet itself, which occurred concurrently with the halo expansion dynamics studied above.
Again, we focused on the pinned, constant radius regime (stages ii) and iii) in Figure \ref{fig : ImageSequence}).
We used images from the same series of experiments used for measuring halo dynamics previously, with additional measurements at 70\%RH and 80\%RH.
With geometrical analysis of optical reflections of the illumination ring light (see \emph{Materials and Methods} and \emph{Supporting Information}), we extracted the temporal evolution of the volume, $V$, of the droplet.
We performed this analysis until spontaneous depinning of the contact line.

\begin{figure}[]
	\centering
	\includegraphics[scale=0.9]{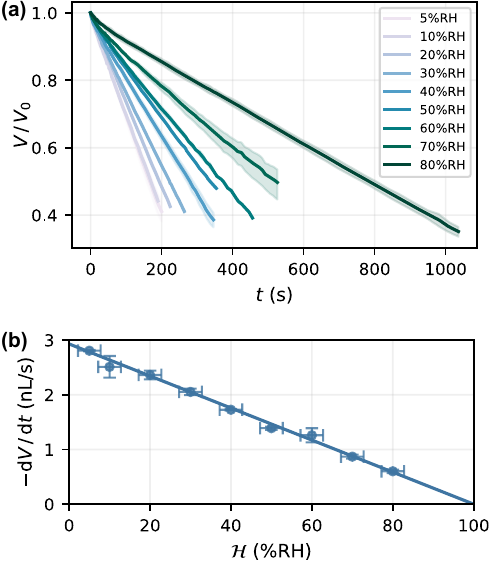}
    \caption{
		\small
		Evaporation dynamics of the pinned droplet as a function of the imposed RH.
		(a) Measured temporal evolution of the volume, $V$, of the droplet normalized by its initial value ($V_0 \simeq \qty{1}{\micro\liter}$ with some fluctuations for each experiment).
		Similarly to Figure \ref{fig : Halo_Dynamics}a, the data is averaged over at least 5 experiments for each RH, and the shaded areas indicate the observed variations across these experiments.
		(b) Extracted steady-state volumetric flow rates.
		The continuous blue line is a fit using Equation \ref{eq : FitFormula_DropletEvapRate}.
	}
	\label{fig : Drop_All}
\end{figure}

Figure \ref{fig : Drop_All}a presents the results at different RH, showing average behavior and standard deviation across at least 5 experimental repeats at each RH.
For clarity, we have normalized in Figure \ref{fig : Drop_All}a the droplet volume by the initial volume, $V_0 = V(t=0)$ in order to compensate for natural fluctuations in $V_0$ coming from the manual drop deposition process ($V_0 = \qty{0.95(0.08)}{\micro\liter}$ across all experiments).

The data in Figure \ref{fig : Drop_All}a shows a short initial transient, followed by a linear decrease of the droplet volume as a function of time.
The early transient cannot be explained by vertical imbibition of the droplet in the porous layer underneath it, which should occur within milliseconds; most likely, this unsteady situation reflects the initial growth of the imbibition halo prior to steady-state (Figure \ref{fig : Halo_Dynamics}), see \emph{Supporting Information}.

The following linear decrease of $V(t)$ in Figure \ref{fig : Drop_All}a suggests a constant evaporation rate.
While this may seem surprising, because the droplet shape is constantly evolving (decrease of contact angle, $\theta$ over time), such an effect is compatible with Equation \ref{eq : DropletFlux}, which predicts that the mass flow rate of evaporation of a sessile droplet only depends on the droplet base radius $\idx{R}{drop}$ when $\theta$ is sufficiently small ($<\qty{40}{\degree}$).
In our experiments, the droplet had a constant value of $\idx{R}{drop}$ during the pinned regime, and $\theta < \qty{30}{\degree}$ (see \emph{Materials and Methods}), which explains the apparent steady-state evaporation regime at large times in Figure \ref{fig : Drop_All}a.

We extracted the slope, $- \odv{V}/{t}$, in the linear regime for all series of experiments at different RH, see Figure \ref{fig : Drop_All}b.
Then, we compared $- \rho \odv{V}/{t}$ to the expected diffusive evaporation mass flow rate for a sessile droplet of the same volume and shape on a flat, nonporous surface, resulting in
\begin{equation}
	\label{eq : FitFormula_DropletEvapRate}
	- \odv{V}{t} = 4 \idx{R}{drop} \epsilon D \left( 1 - \rh \right)
\end{equation}
from Equations \ref{eq : DropletFlux} and \ref{eq : DeltaH_Drop}.
We fitted the data in Figure \ref{fig : Drop_All}b with Equation \ref{eq : FitFormula_DropletEvapRate}, keeping $\idx{R}{drop}$ as the only fitting parameter ($\epsilon$ and $D$ being tabulated water properties, see \emph{Supporting Information}).
This procedure yielded $\idx{R}{drop} = \qty{1.30(0.03)}{\mm}$, in near perfect agreement with $\idx{R}{drop} = \qty{1.34(0.04)}{\mm}$ obtained from image analysis.
This correspondence suggests that volumetric variations of the droplets are dominated by direct, diffusive evaporation into the surrounding air and that the imbibition flow into the pores associated with halo formation has negligible impact on that volumetric dynamics.

\subsection{Evaporation rates}

Since variations in volume (Figure \ref{fig : Drop_All}b) reflect evaporation rates from the drop, and since transient halo development times $\tau(\rh)$ (Figure \ref{fig : Halo_FitResults}a) directly relates to evaporation fluxes from the nanopores (Equation \ref{eq : tau}), our experimental data enables an independent estimation of the evaporation rates of bulk and confined water.

\begin{figure}[]
	\centering
	\includegraphics[scale=0.9]{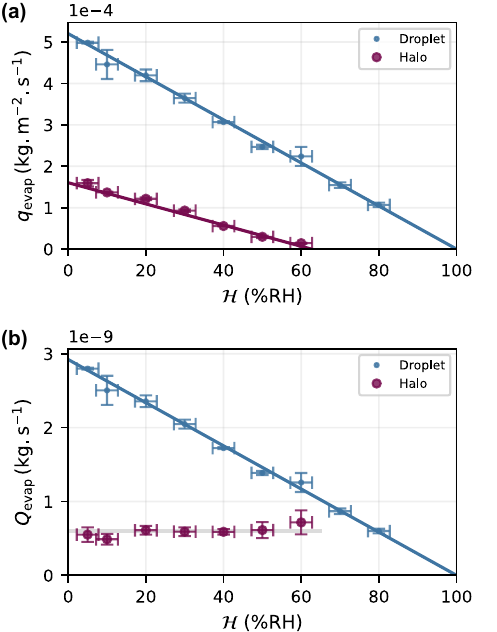}
    \caption{
		\small
		Comparison of evaporation rates between the droplet (bulk water) and the halo (confined water), as a function of imposed RH.
		(a) Evaporation flux (mass flow rate by unit area) and (b) corresponding total mass flow rates at steady-state.
		Droplet data is calculated from changes in drop volume (Figure \ref{fig : Drop_All}b), while halo data is extracted from the transient evaporation times (Figure \ref{fig : Halo_FitResults}a), see text for details.
		The continuous red line in (a) represents Equations \ref{eq : tau}, \ref{eq : FitFormula_Tau} and \ref{eq : FitFormula_Tau0}.
		The continuous blue lines in (a) and (b) correspond to the fitting line in Figure \ref{fig : Drop_All}b, i.e.,
		Equation \ref{eq : FitFormula_DropletEvapRate}.
		The gray line in (b) is a guide for the eye, showing a constant value for $Q_\mathrm{evap}$.
	}
	\label{fig : Evap_All}
\end{figure}

Figure \ref{fig : Evap_All} shows the extracted steady-state evaporation rates per unit area (mass flux, $q_\mathrm{evap}$, Figure \ref{fig : Evap_All}a) and the corresponding total mass flow rates ($Q_\mathrm{evap}$, Figure \ref{fig : Evap_All}).
For the droplet data, we first calculated $\idxy{Q}{evap}{drop} = - \rho \odv{V}/{t}$, using the data in Figure \ref{fig : Drop_All}b and then $\idxy{q}{evap}{drop}$ from Equation \ref{eq : qdrop}.
For halo evaporation, we calculated $\idxy{q}{evap}{halo}$ from $\tau(\rh)$ (Figure \ref{fig : Halo_FitResults}a) using Equation \ref{eq : tau}, then estimated $\idxy{Q}{evap}{halo} = \mathcal{A}_\mathrm{halo} \times q_\mathrm{evap}^\mathrm{(halo)}$, with $\mathcal{A}_\mathrm{halo} = \pi \left[ (\idx{R}{drop} + \idx{L}{max})^2 - \idx{R}{drop}^2 \right]$ the steady-state area of the halo calculated from its maximal extension ($\idx{L}{max}$, data in Figure \ref{fig : Halo_FitResults}c).

The cuves in Figure \ref{fig : Evap_All}a show two major distinctions between droplet and halo evaporation fluxes.
First, the external RH at which evaporation vanishes is different: 100\%RH for the droplet, as is expected for the equilibrium of bulk water, and $\idx{\rh}{eq} \simeq 63$\%RH for the pore liquid in the halo, due tu confinement-induced reduction in equilibrium RH (Kelvin effect, see \emph{Materials and Methods}).
This marked difference is a clear illustration of how the Kelvin effect (a thermodynamic concept) may strongly impact transport dynamics, as has been recently shown in other contexts such as fluid pervaporation \cite{Vincent2016,Lei2016} or condensation-driven imbibition flows \cite{Vincent2017,Zhong2018}.

A second distinction visible in Figure \ref{fig : Evap_All}a is the difference in magnitude between the droplet and halo evaporation fluxes.
This difference is also visible in the values of effective boundary layer thicknesses, $\delta = 2.2$ mm calculated for halo evaporation (see above) and $\delta_\mathrm{drop} = 1.0$ mm for drop evaporation, calculated using Equation \ref{eq : EffectiveDeltaDrop}.
As discussed in \emph{Theory}, diffusive evaporation of droplets occurs in a three-dimensional fashion with characteristic distances not set by external conditions but by the drop geometry;
$\delta_\mathrm{drop}$ is thus an equivalent boundary layer thickness, which represents 3D droplet diffusion with 1D boundary layer concepts.
On the contrary, halo evaporation is dictated by actual boundary layers in the system (see \emph{Theory}).
The difference between $\delta$ and $\delta_\mathrm{drop}$ is thus not a disagreement, and in fact it might rather be a coincidence that they are close in magnitude.


In terms of total evaporation flow rates ($\idx{Q}{evap}$), the data in Figure \ref{fig : Evap_All}b shows that the steady-state halo evaporation rate $\idxy{Q}{evap}{halo}$ does not significantly depend on RH.
This interesting independence suggests that when RH increases, the decrease of the evaporation rate by unit area, $q_\mathrm{evap}$, is balanced by the increase in halo area, resulting in a constant product $\mathcal{A}_\mathrm{halo} \times q_\mathrm{evap}^\mathrm{(halo)}$.
While we have a good understanding of the physics describing the reduction in $q_\mathrm{evap}$ -- boundary layer diffusion and Kelvin effect (Equations \ref{eq : BoundaryLayerFlux}--\ref{eq : DeltaH_Halo}) -- predicting changes in steady-state halo dimensions is more challenging.
Indeed, calculating $\idx{L}{max} \sim \sqrt{w \tau}$ requires knowledge of $w(\rh)$, for which we do not have predictive theory at the moment, but only empirical fit formula (see \emph{Halo expansion dynamics} above and \emph{Lateral vapor transport} below).
It is thus not obvious whether the constant value of $\idxy{Q}{evap}{halo}$ in Figure \ref{fig : Evap_All}b has fundamental reasons or is a mere coincidence.

Due to the dynamic equilibrium between capillary flow and evaporation, $\idxy{Q}{evap}{halo}$ should also be equal to the imbibition flow rate in the halo, $\idx{Q}{cap}$.
This statement however raises two contradictions.
First, if these rates were equal, Figure \ref{fig : Evap_All}b would suggest that $\idx{Q}{cap} = \idxy{Q}{evap}{halo}$ would not be negligible compared to the drop evaporation rate $\idxy{Q}{evap}{drop}$, especially at large RH.
If this was the case, capillary pumping should account for a significant part of the observed volumetric variations of the droplet, contradicting the conclusion of the analysis of droplet evaporation dynamics made above.
Second, the values of $\idxy{Q}{evap}{halo}$ do not match the expected order of magnitude for $\idx{Q}{cap}$.
Indeed, using Equations \ref{eq : LW} and \ref{eq : Darcy}
\begin{equation}
	\label{eq : LW_Flow_Estimate}
	Q_\mathrm{cap} \simeq \pi \phi \idx{R}{drop} H \rho w / \idx{L}{max}
\end{equation}
where we have assumed that the pressure gradient driving the flow is $\nabla P \simeq \DeltaPc / \idx{L}{max}$, through a cross-section area $2 \pi \idx{R}{drop} H$.
For $\idx{L}{max} \simeq 1$ mm and $w = w_0 = 7 \times 10^{-8} \, \mathrm{m^2/s}$ based on material parameters (see \emph{Supporting Information}) Equation \ref{eq : LW_Flow_Estimate} predicts $Q_\mathrm{cap} \simeq 5 \times 10^{-11} \, \mathrm{kg/s}$, i.e. one order of magnitude smaller than the values shown in Figure \ref{fig : Evap_All}b for $\idxy{Q}{evap}{halo}$.
In the next sections, we evaluate whether the apparent contradictions raised above can be solved by considering alternative mechanisms of transport.

\subsection{Effect of pore pre-filling}

Here, we evaluate whether partial pore filling prior to droplet deposition can explain the unexpected dependence on RH of the Lucas-Washburn coefficient, $w$, extracted from our experimental data (Figure \ref{fig : Halo_FitResults}b) and the apparent discrepancies in the evaporation rates discussed in the previous section.

To that aim, it is useful to inspect Equation \ref{eq : LW}, which relates $w$ to 3 material parameters.
The permeability, $\kappa$, accounts for the viscous resistance to liquid flow inside the wetted zone of the porous medium situated between the droplet and the imbibition front; there is no reason to think that $\kappa$ should be impacted by the external RH because it describes transport in a part of the porous layer that is always full with liquid.
On the other hand, the porosity, $\phi$ describes the initially empty volume in the pores that gets filled due to the imbibition process.
As a result, if pores are partially filled (filling fraction, $f$) initially, e.g. by vapor adsorption or condensation, $\phi$ changes.
Pre-filling also potentially impacts capillary pressure, $\DeltaPc$, by modifying the shape of the advancing meniscus in the pore space.

\begin{figure}[]
	\centering
	\includegraphics[scale=0.9]{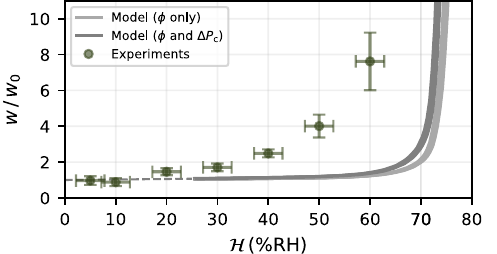}
    \caption{
		\small
		Prediction of the increase in the lateral transport parameter (LW coefficient), $w$, based on the amount of pre-filling of the pores measured from water sorption isotherms (Figure \ref{fig : Isotherm}).
		The circles correspond to the experimental data as in Figure \ref{fig : Halo_FitResults}b.
		The light gray line is a prediction based on the variation of the effective porosity of the sample as a function of RH (Equation \ref{eq : Expected_W_phi}), while the darker gray data is a prediction that combines the effect of filling fraction on both the effective porosity and the capillary pressure (Equation \ref{eq : Expected_W_phi_Pc}).
		The dashed, gray line corresponds to an extrapolation of the models to 0\%RH.
	}
	\label{fig : W_Increase}
\end{figure}

Let's first assume that only porosity is modified by pre-filling.
Adsorption or condensation of water vapor result in $f > 0$ (see Figure \ref{fig : Isotherm}), resulting in a porosity decrease by a factor $1 - f$.
Therefore using Equation \ref{eq : LW} we would expect the LW coefficient to follow
\begin{equation}
	\label{eq : Expected_W_phi}
	\frac{w}{w_0} = \frac{1}{1 - f}
\end{equation}
where $w_0$ corresponds to a fully dry case ($f=0$, obtained for a relative humidity, $\rh=0$).
In Figure \ref{fig : W_Increase} (light gray), we compare experimental data (Figure \ref{fig : Halo_FitResults}b, normalized by $w_0$ obtained from fitting with Equation \ref{eq : FitFormula_w}) with the prediction of Equation \ref{eq : Expected_W_phi}, using water sorption isotherms (Figure \ref{fig : Isotherm}) to evaluate $f(\rh)$.
While Equation \ref{eq : Expected_W_phi} predicts a strong increase of $w$ as observed experimentally, this increase should be much more localized around the capillary condensation transition ($\simeq 70$\%RH), and with a much more abrupt increase compared to the progressive increase of $w / w_0$ seen in experiments.

With our definition of the front position, capillary pressure at the halo edge should not be impacted by pre-filling, and its value should stay in the vicinity of $\left( \DeltaPc \right)_\mathrm{eq}$ due to local thermodynamic equilibrium (see \emph{Materials and Methods} and \emph{Theory}).
Here, we temporarily ignore this fact, and estimate how much pre-filling could impact capillary pressure based on a naive geometrical approach.
Considering cylindrical pores (geometrical radius, $r_\mathrm{p}$) with an adsorbed film of water of thickness, $h$, on the pore walls, the empty part of the pore is a cylinder of radius, $r = r_\mathrm{p} - h$, and corresponds to a filling fraction, $f = 1 - \left( r/r_\mathrm{p} \right)^2$.
Since from Laplace's law, the capillary pressure, $\DeltaPc$ scales inversely with the radius of curvature of the meniscus, we assume that $\DeltaPc / (\DeltaPc)_0 \simeq r_\mathrm{p} / r$, which equals $(1 - f)^{-1/2}$ using the previous formula relating $f$ to $r$.
Combining this scaling with the one already established for $\phi(f)$ leading to Equation \ref{eq : Expected_W_phi}, we finally obtain
\begin{equation}
	\label{eq : Expected_W_phi_Pc}
	\frac{w}{w_0} \simeq \frac{1}{\left( 1 - f \right)^{3/2}}
\end{equation}
which also has a prediction very different from experimental data (Figure \ref{fig : W_Increase}, dark gray).

The analysis above indicates that pre-filling of the pores is not sufficient to explain the observed increase of $w$ with RH, because increases in filling fraction prior to capillary condensation are not significant enough to produce large changes in effective porosity or capillary pressure.

Potentially, the external, partially filled zone around the halo (see \emph{Materials and Methods}) might also impact $w$ by reducing the effective porosity.
However, as discussed previously, if the molecules required to fill this external zone originate from the halo liquid itself, we would rather expect a slowing down of the front rather than an acceleration (see \emph{General considerations} in \emph{Theory}).
More generally, any porosity-induced pre-filling effect, whether from initial adsorption or from a partially filled zone that develops around the halo, would impact only transients, because at steady-state no additional porosity is filled by the stationary imbibition front.
On contrary, our experiments show a consistently increased $w$ that manifests itself both in the initial transient dynamics $L(t) \simeq \sqrt{w t}$ and in the steady-state halo extension $\idx{L}{max} \simeq \sqrt{w \tau}$ (see \emph{Theory} and Figure \ref{fig : Halo_Dynamics}).
We thus need an alternative explanation to pre-filling for the unexpected magnitude and variations of $w(\rh)$.
In the next section, we evaluate the role of vapor flow outside the pores, in parallel with the capillary flow within the pores.

\subsection{Lateral vapor transport}

\begin{figure}
	\centering
	\includegraphics[scale=0.9]{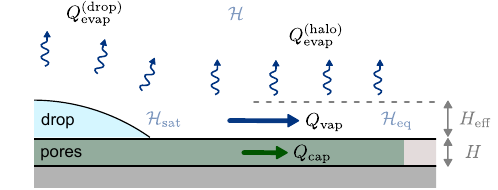}
    \caption{
		\small
		Schematic description of hypothetic water mass flow rates in the system, including a vapor transport of magnitude $\idx{Q}{vap}$, in parallel with the liquid capillary flow, $\idx{Q}{cap}$.
		Color coding follows that of Figure \ref{fig : ExpSetup}.
	}
	\label{fig : VaporCondensation}
\end{figure}

Here, we propose a potential solution to the apparent contradictions in both evaporation rates and lateral transport coefficients raised in the previous sections.
We hypothesize that the liquid present in the halo does not only come from capillary pumping, but also from a fraction of the vapor molecules evaporated from the droplet surface, transported laterally along the sample surface, and condensed in the pores (see Figure \ref{fig : VaporCondensation}).
Such transport is thermodynamically possible and favorable because owing to the Kelvin effect, the equilibrium RH of the liquid in the droplet, $\rh_\mathrm{sat} = 1$, is larger than that of the liquid in the pores, $\idx{\rh}{eq} < 1$, resulting in a driving force $\idx{\rh}{sat} - \idx{\rh}{eq} > 0$.
Also, because of the low liquid permeability of mesoporous media, vapor transport can become comparable or even more efficient than capillary flow, depending on the geometry in which vapor transport occurs \cite{Vincent2024}.

This hypothetical condensation flow (of magnitude $\idx{Q}{vap}$) superimposes on the global evaporation of liquid from the halo to the far field, $\idxy{Q}{evap}{halo}$, driven by $\idx{\rh}{eq} - \rh > 0$.
As a result, the net mass flow rate of evaporation from the halo is now $\idxy{Q}{evap}{halo} - \idx{Q}{vap}$.
This separation between $\idxy{Q}{evap}{halo}$ and $\idx{Q}{vap}$ is somewhat artificial and corresponds in reality to a single diffusion flux coupled to a three-dimensional water vapor pressure field in the air surrounding the droplet and the porous surface.
However decomposing in such a way allows us to make simple arguments in terms of expected behavior and orders of magnitude, see below.

We assume that $\idx{Q}{vap}$ is part of the total droplet evaporation flow rate $\idxy{Q}{evap}{drop}$ and constitutes a fraction of it.
As a result mass conservation for the droplet simply writes
\begin{equation}
	\label{eq : Qdrop}
	\idxy{Q}{evap}{drop} = \idx{Q}{drop} - \idx{Q}{cap}
\end{equation}
where $\idx{Q}{drop} = -\rho \odv{V}/{t}$ is the total mass flow rate lost by the droplet under the combined effect of evaporation and capillary flow.
On the other hand, the dynamic equilibrium in the halo now equates the global halo (vertical) evaporation rate, $\idxy{Q}{evap}{halo}$, to the sum of all lateral transport:
\begin{equation}
	\label{eq : Qhalo}
	\idxy{Q}{evap}{halo} = \idx{Q}{cap} + \idx{Q}{vap}
\end{equation}
If $\idx{Q}{cap}$ is small compared to evaporation rates as estimated above (see Equation \ref{eq : LW_Flow_Estimate}), then $\idxy{Q}{evap}{drop} \simeq \idx{Q}{drop}$ and $\idxy{Q}{evap}{halo} \simeq \idx{Q}{vap}$, which results in decorrelation between $\idxy{Q}{evap}{drop}$ and $\idxy{Q}{evap}{halo}$.
This decoupling solves the apparent contradiction raised in the \emph{Evaporation rates} section above, because it explains why the halo evaporation rate can be non-negligible compared to the droplet evaporation rate in Figure \ref{fig : Evap_All}b, while at the same time not influencing noticeably droplet evaporation.

We also define a dimensionless factor $\beta = \idx{Q}{vap} / \idx{Q}{cap}$, which compares the relative magnitudes of lateral vapor transport by condensation and capillary flow.
The dynamic equilibrium (Equation \ref{eq : Qhalo}) thus rewrites $\idxy{Q}{evap}{halo} = \idx{Q}{cap} \left( 1 + \beta \right)$, resulting in an increase in the apparent transport coefficient
\begin{equation}
	\label{eq : ww0cond}
	w = w_0 \left( 1 + \beta \right)
\end{equation}
compared to the case without lateral vapor transport ($\idxy{Q}{evap}{halo} = \idx{Q}{cap}$ where $\idx{Q}{cap} \propto w_0$).
Recently, it was shown that vapor evaporation, diffusion, and condensation in air-filled microchannels (mass flux $\idx{q}{vap}$) coupled via the Kelvin equation to a mesoporous medium with capillary flow ($\idx{q}{cap}$) resulted in $\idx{q}{vap} / \idx{q}{cap} = \epsilon^2 \langle \rh \rangle D / (\psat \kappa)$, with $q$ the mass flux and $\langle \rh \rangle$ the average humidity in the channel \cite{Vincent2024}.
In our system, the vapor flow is not contained in a channel, but we assume that it occurs on a typical, effective height, $\idx{H}{eff}$, so that $\idx{H}{eff} \idx{q}{vap} / (H \idx{q}{cap})$ is an estimate of the ratio of flow rates, $\idx{Q}{vap} / \idx{Q}{cap} = \beta$.
As a result,
\begin{equation}
	\label{eq : beta}
	\beta \simeq \epsilon^2 \langle \rh \rangle \left( \frac{\idx{H}{eff}}{H} \right) \frac{D}{\psat \kappa}
\end{equation}
where $\langle \rh \rangle$ represents an average RH in the zone between the droplet and the halo where vapor transport occurs.
Equation \ref{eq : beta} hints as to why the apparent imbibition coefficient, $w$ (Equation \ref{eq : ww0cond}) could be RH-dependent, although it is not obvious why $\langle \rh \rangle$ would be strongly impacted by the RH imposed far away from the droplet, $\rh$.
Also, Equations \ref{eq : ww0cond}-\ref{eq : beta} cannot explain the apparent divergence of $w(\rh)$ seen in our data (Figure \ref{fig : W_Increase}).
Nonetheless, in terms of orders of magnitude, these equations are compatible with observed increases in $w(\rh)$ ($\beta \simeq 8$ at $\rh = 0.6$) if one uses $\idx{H}{eff} \simeq \qty{0.3}{\mm}$.
This value is comparable to our system's dimensions, e.g., the initial height of the droplet after deposition is $h = \idx{R}{drop} \tan (\theta / 2) \simeq \qty{0.36}{\mm}$.

Equation \ref{eq : beta} also suggests that it may be possible to tune the relative importance of condensation and capillary flows by changing the architecture and geometry of the porous medium (e.g., thickness or permeability), or the fluid properties ($\epsilon$, $\psat$), which opens perspectives for experimental suppression of magnification of the effect.
Published literature about halo dynamics (e.g., \cite{Mercuri2017,Pizarro2024,Czerwenka2025}) often involves much thinner mesoporous films ($H \sim \qtyrange{0.2}{0.6}{\um}$) but with similar porosities and larger pore diameters ($\sim \qtyrange{4}{12}{\nm}$) than our poSi layers ($H \simeq \qty{5}{\um}$, diameter $\simeq \qtyrange{3}{4}{\nm}$), resulting in larger $\kappa \propto \idx{r}{p}^2$ \cite{Vincent2016}.
Therefore, from Equation \ref{eq : beta}, these studies had potentially similar or larger $\beta$ than in our study, meaning that they could have been in a regime dominated by vapor flow instead of capillary flow.

We conclude that lateral vapor transport originating from droplet evaporation and condensation into the halo is a plausible process occurring in our system and others.
In our work, taking this effect into account solves apparent contradictions in evaporation rates and transport coefficients.
While some studies already mentioned vapor condensation effects and questioned their role in droplet infiltration dynamics \cite{Ceratti2015,Urteaga2019,Khalil2020a,Martinez2024,Hartmann2025}, we provide for the first time potential scaling laws and orders of magnitude governing the balance between vapor condensation and capillary flow, even if more work is required to theoretically predict transport rates and their dependence on RH.
Numerical modeling \cite{Hartmann2025} and spatially resolved measurements of pore filling around droplets \cite{Sallese2020,Martinez2024} are promising tools to further characterize this complex dynamics.

\section{Conclusion}

In this study, we presented a detailed analysis of water droplet infiltration and evaporation after deposition on a thin mesoporous layer, with particular emphasis on the dynamics of the wetted annulus (halo) that develops in the pores around the droplet.
We used a combined theoretical and experimental approach to systematically investigate how the relative humidity of the air impacted halo dynamics and droplet evaporation.
From our analysis of transients, we extracted independently evaporation rates from the bulk droplet fluid and from the liquid confined in the pores, showing that the latter was strongly impacted by nanoconfinement (Kelvin effect), and providing a remarkable illustration of how nanoscale behavior of fluids can be inferred from macroscopic measurements.

We also extracted a radial transport parameter, $w$, which exhibited a strong and unexpected dependence on RH. We could not explain this behavior solely from material parameters, suggesting that capillary flow within the pores is not the only process resulting in radial transport from the droplet to the porous medium.
As such, the parameter $w$ that we have extracted is an effective transport coefficient, which includes both capillary flow (Lucas-Washburn) and other processes that occur in parallel.
We have shown that a likely mechanism is a flow of water vapor in the air along the porous surface, driven by the differences in equilibrium RH between the bulk fluid in the droplet and the confined liquid in the pores.
Such a process is made possible, again, by the Kelvin effect.
We have shown that the relative importance of this parallel flow can in principle be enhanced or suppressed when changing the internal architecture (pore size, porosity, etc.) and macroscopic geometry (thickness) of the material.
These results such suggest experimental tests of the predicted scaling laws, as well as theoretical and numerical work to understand the complex coupling between capillary flow, phase change, vapor transport and confined thermodynamics.

Practically, our study also shows that it is possible to finely tune the invasion patterns of fluid droplets in mesoporous materials by playing on the external RH only.
For example, we were able to increase the halo width by an order of magnitude, from hundreds of micrometers to several millimeters by raising the humidity from \qty{5}{\percent}RH to \qty{60}{\percent}RH.
But halos of arbitrary large sizes can be obtained due to a divergence caused by the Kelvin effect, at the expense of the dynamics slowing down with increasing RH.
The critical RH for that divergence is controllable through the pore size and wettability of the material.
The wide range of patterns achievable using a simple control parameter such as RH is potentially interesting in various contexts such as RH-based sensing and actuation, or applications involving precise and selectable fluid deposition in porous materials such as printing or drug delivery, e.g., in plant tissues.

More fundamentally, our results indicate that various important effects need to be taken into account when predicting or analyzing droplet infiltration data: e.g., imbibition/evaporation flux coupling introduces a factor 2 in the analysis of transient times (and $\sqrt{2}$ for halo extensions); 2D effects become important for large halo extensions; capillary transport can become negligible in the total lateral transport mechanisms, in favor of evaporation/condensation flows; response of the system is highly RH-dependent with divergences due to the Kelvin effect, etc.
Taking these effects into consideration is crucial for the interpretation of droplet infiltration data, with particular importance when using such systems as platforms for studying properties of confined fluids, or for applications requiring precise knowledge of deposition patterns from droplet imbibition.

\subsection*{Acknowledgements}

The authors thank G. Simon and F. Gay for technical support in mechanics and electronics, respectively, and S. Le Floch for help with sample preparation.
This work was supported by the French National Research Agency (ANR) under grant ANR-19-CE09-0010 (SINCS) and by the European Union through the EHAWEDRY project (H2020-FETOPEN, Grant Agreement No. 964524).

\subsection*{Data availability statement}

Data underlying this study is openly available in the Zenodo repository at
https://doi.org/10.5281/zenodo.18019169.

\subsection*{Supporting Information}

Details on sample characterization, image analysis methods, numerical values for the properties of water, other numerical estimates (LW coefficient, time scales of vertical imbibition, droplet initial transient), and misc. discussion (time origin in the fitting procedure, equilibrium RH in the halo vs. estimated from isotherms).

\bibliography{biblio}

@article{Al-Samahiji2000,
  title = {Degree and {{Extent}} of {{Wetting Due}} to {{Capillary Rise}} in {{Soils}}},
  author = {{Al-Samahiji}, Despina and Houston, Sandra L. and Houston, William N.},
  year = 2000,
  month = jan,
  journal = {Transportation Research Record: Journal of the Transportation Research Board},
  volume = {1709},
  number = {1},
  pages = {114--120},
  issn = {0361-1981, 2169-4052},
  doi = {10.3141/1709-14},
  urldate = {2025-10-02},
  copyright = {https://journals.sagepub.com/page/policies/text-and-data-mining-license},
  langid = {english}
}

@article{Aslannejad2021,
  title = {Liquid Droplet Imbibition into a Thin Coating Layer: {{Direct}} Pore-Scale Modeling and Experimental Observations},
  shorttitle = {Liquid Droplet Imbibition into a Thin Coating Layer},
  author = {Aslannejad, H. and Loginov, S.V. and Van Der Hoek, B. and Schoonderwoerd, E.M. and Gerritsen, H.C. and Hassanizadeh, S.M.},
  year = 2021,
  month = feb,
  journal = {Progress in Organic Coatings},
  volume = {151},
  pages = {106054},
  issn = {03009440},
  doi = {10.1016/j.porgcoat.2020.106054},
  urldate = {2025-10-02},
  langid = {english}
}

@article{Bell1906,
  title = {The {{Flow}} of {{Liquids}} through {{Capillary Spaces}}},
  author = {Bell, J. M. and Cameron, F. K.},
  year = 1906,
  month = nov,
  journal = {The Journal of Physical Chemistry},
  volume = {10},
  number = {8},
  pages = {658--674},
  issn = {0092-7325, 1541-5740},
  doi = {10.1021/j150080a005},
  urldate = {2025-10-02},
  langid = {english}
}

@article{Bellezza2025,
  title = {Salt Crystallization and Deliquescence Triggered by Humidity Cycles in Nanopores},
  author = {Bellezza, Hugo and Poizat, Marine and Vincent, Olivier},
  year = 2025,
  journal = {arXiv preprint},
  pages = {arXiv:2510.27309},
  doi = {10.48550/ARXIV.2510.27309},
  urldate = {2025-11-17},
  copyright = {arXiv.org perpetual, non-exclusive license}
}

@inproceedings{Binder2010,
  title = {Test {{Method To Quantify The Wicking Properties Of Porous Insulation Materials Designed To Prevent Interstitial Condensation}}},
  booktitle = {Porous {{Media And Its Applications In Science}}, {{Engineering}}, {{And Industry}}: 3rd {{International Conference}}},
  author = {Binder, Andrea and Zirkelbach, Daniel and K{\"u}nzel, Hartwig and Vafai, Kambiz},
  year = 2010,
  pages = {242--247},
  address = {Montecatini (Italy)},
  doi = {10.1063/1.3453818},
  urldate = {2025-10-02}
}

@article{Bossert2020,
  title = {Stress or {{Strain Does Not Impact Sorption}} in {{Stiff Mesoporous Materials}}},
  author = {Bossert, M. and Grosman, A. and Trimaille, I. and No{\^u}s, C. and Rolley, E.},
  year = 2020,
  month = sep,
  journal = {Langmuir},
  volume = {36},
  number = {37},
  pages = {11054--11060},
  issn = {0743-7463, 1520-5827},
  doi = {10.1021/acs.langmuir.0c01939},
  urldate = {2025-08-09},
  copyright = {https://doi.org/10.15223/policy-029},
  langid = {english}
}

@article{Brutin2018,
  title = {Recent Advances in Droplet Wetting and Evaporation},
  author = {Brutin, D. and Starov, V.},
  year = 2018,
  journal = {Chemical Society Reviews},
  volume = {47},
  number = {2},
  pages = {558--585},
  issn = {0306-0012, 1460-4744},
  doi = {10.1039/C6CS00902F},
  urldate = {2023-12-15},
  langid = {english}
}

@article{Casanova2012,
  title = {Controlling the {{Role}} of {{Nanopore Morphology}} in {{Capillary Condensation}}},
  author = {Casanova, F{\`e}lix and Chiang, Casey E. and Ruminski, Anne M. and Sailor, Michael J. and Schuller, Ivan K.},
  year = 2012,
  month = may,
  journal = {Langmuir},
  volume = {28},
  number = {17},
  pages = {6832--6838},
  issn = {0743-7463, 1520-5827},
  doi = {10.1021/la204933m},
  urldate = {2019-07-05},
  langid = {english}
}

@article{Caupin2008,
  title = {Absolute Limit for the Capillary Rise of a Fluid},
  author = {Caupin, Fr{\'e}d{\'e}ric and Cole, Milton W. and Balibar, S{\'e}bastien and Treiner, Jacques},
  year = 2008,
  month = jun,
  journal = {EPL (Europhysics Letters)},
  volume = {82},
  number = {5},
  pages = {56004},
  issn = {0295-5075, 1286-4854},
  doi = {10.1209/0295-5075/82/56004},
  urldate = {2019-07-05},
  langid = {english}
}

@article{Ceratti2015,
  title = {Critical Effect of Pore Characteristics on Capillary Infiltration in Mesoporous Films},
  author = {Ceratti, D. R. and Faustini, M. and Sinturel, C. and Vayer, M. and Dahirel, V. and Jardat, M. and Grosso, D.},
  year = 2015,
  journal = {Nanoscale},
  volume = {7},
  number = {12},
  pages = {5371--5382},
  issn = {2040-3364, 2040-3372},
  doi = {10.1039/C4NR03021D},
  urldate = {2019-09-22},
  langid = {english}
}

@article{Chabas2000,
  title = {Crystallization and Dissolution of Airborne Sea-Salts on Weathered Marble in a Coastal Environment at {{Delos}} ({{Cyclades}}--{{Greece}})},
  author = {Chabas, A and Jeannette, D and Lef{\`e}vre, R.A},
  year = 2000,
  month = jan,
  journal = {Atmospheric Environment},
  volume = {34},
  number = {2},
  pages = {219--224},
  issn = {13522310},
  doi = {10.1016/S1352-2310(99)00256-3},
  urldate = {2025-10-02},
  copyright = {https://www.elsevier.com/tdm/userlicense/1.0/},
  langid = {english}
}

@article{Czerwenka2025,
  title = {Self-{{Driven Fluid Imbibition}} of {{Salt Solutions}} into {{Mesoporous Films}}},
  author = {Czerwenka, Laura and {Andrieu-Brunsen}, Annette},
  year = 2025,
  month = jun,
  journal = {Langmuir},
  volume = {41},
  number = {22},
  pages = {13892--13901},
  issn = {0743-7463, 1520-5827},
  doi = {10.1021/acs.langmuir.5c00650},
  urldate = {2025-07-08},
  copyright = {https://doi.org/10.15223/policy-029},
  langid = {english}
}

@article{Davis1999,
  title = {Spreading and Imbibition of Viscous Liquid on a Porous Base},
  author = {Davis, S. H. and Hocking, L. M.},
  year = 1999,
  month = jan,
  journal = {Physics of Fluids},
  volume = {11},
  number = {1},
  pages = {48--57},
  issn = {1070-6631, 1089-7666},
  doi = {10.1063/1.869901},
  urldate = {2025-08-06},
  langid = {english}
}

@article{Dimitrov2007,
  title = {Capillary {{Rise}} in {{Nanopores}}: {{Molecular Dynamics Evidence}} for the {{Lucas-Washburn Equation}}},
  shorttitle = {Capillary {{Rise}} in {{Nanopores}}},
  author = {Dimitrov, D. I. and Milchev, A. and Binder, K.},
  year = 2007,
  month = jul,
  journal = {Physical Review Letters},
  volume = {99},
  number = {5},
  pages = {054501},
  publisher = {American Physical Society},
  doi = {10.1103/PhysRevLett.99.054501},
  urldate = {2020-12-28}
}

@article{Eddi2013,
  title = {Short Time Dynamics of Viscous Drop Spreading},
  author = {Eddi, A. and Winkels, K. G. and Snoeijer, J. H.},
  year = 2013,
  month = jan,
  journal = {Physics of Fluids},
  volume = {25},
  number = {1},
  pages = {013102},
  issn = {1070-6631, 1089-7666},
  doi = {10.1063/1.4788693},
  urldate = {2025-08-06},
  langid = {english}
}

@article{Elliott2021,
  title = {Surface Thermodynamics at the Nanoscale},
  author = {Elliott, Janet A. W.},
  year = 2021,
  month = may,
  journal = {The Journal of Chemical Physics},
  volume = {154},
  number = {19},
  pages = {190901},
  issn = {0021-9606, 1089-7690},
  doi = {10.1063/5.0049031},
  urldate = {2023-05-11},
  langid = {english}
}

@article{Fischer2022,
  title = {Wicking Dynamics in Yarns},
  author = {Fischer, Robert and Schlep{\"u}tz, Christian M. and Zhao, Jianlin and Boillat, Pierre and Hegemann, Dirk and Rossi, Ren{\'e} M. and Derome, Dominique and Carmeliet, Jan},
  year = 2022,
  month = nov,
  journal = {Journal of Colloid and Interface Science},
  volume = {625},
  pages = {1--11},
  issn = {00219797},
  doi = {10.1016/j.jcis.2022.04.060},
  urldate = {2025-10-02},
  langid = {english}
}

@article{Gimenez2018,
  title = {Directional {{Water Collection}} in {{Nanopore Networks}}},
  author = {Gimenez, Rocio and Bellino, Mart{\'i}n Gonzalo and Berli, Claudio Luis Alberto},
  year = 2018,
  month = nov,
  journal = {ACS Omega},
  volume = {3},
  number = {11},
  pages = {16040--16045},
  issn = {2470-1343, 2470-1343},
  doi = {10.1021/acsomega.8b02376},
  urldate = {2025-10-02},
  copyright = {http://pubs.acs.org/page/policy/authorchoice\_termsofuse.html},
  langid = {english}
}

@article{Gimenez2018a,
  title = {Electrical Current Nanogeneration Driven by Spontaneous Nanofluidic Oscillations},
  author = {Gimenez, R. and Mercuri, M. and Berli, C. L. A. and Bellino, M. G.},
  year = 2018,
  journal = {Nanoscale},
  volume = {10},
  number = {7},
  pages = {3144--3147},
  issn = {2040-3364, 2040-3372},
  doi = {10.1039/C8NR00269J},
  urldate = {2025-10-02},
  langid = {english}
}

@article{Gravelle2016,
  title = {Anomalous Capillary Filling and Wettability Reversal in Nanochannels},
  author = {Gravelle, Simon and Ybert, Christophe and Bocquet, Lyd{\'e}ric and Joly, Laurent},
  year = 2016,
  month = mar,
  journal = {Physical Review E},
  volume = {93},
  number = {3},
  issn = {2470-0045, 2470-0053},
  doi = {10.1103/PhysRevE.93.033123},
  urldate = {2019-08-26},
  langid = {english}
}

@article{Gruener2009,
  title = {Capillary Rise of Water in Hydrophilic Nanopores},
  author = {Gruener, Simon and Hofmann, Tommy and Wallacher, Dirk and Kityk, Andriy V. and Huber, Patrick},
  year = 2009,
  month = jun,
  journal = {Physical Review E},
  volume = {79},
  number = {6},
  issn = {1539-3755, 1550-2376},
  doi = {10.1103/PhysRevE.79.067301},
  urldate = {2019-07-05},
  langid = {english}
}

@article{Ha2020,
  title = {Capillarity in {{Soft Porous Solids}}},
  author = {Ha, Jonghyun and Kim, Ho-Young},
  year = 2020,
  month = jan,
  journal = {Annual Review of Fluid Mechanics},
  volume = {52},
  number = {1},
  pages = {263--284},
  issn = {0066-4189, 1545-4479},
  doi = {10.1146/annurev-fluid-010518-040419},
  urldate = {2025-10-02},
  langid = {english}
}

@article{Hartmann2025,
  title = {Gradient Dynamics Model for Drops of Volatile Liquid on a Porous Substrate},
  author = {Hartmann, Simon and Thiele, Uwe},
  year = 2025,
  month = jan,
  journal = {Physical Review Fluids},
  volume = {10},
  number = {1},
  pages = {014003},
  issn = {2469-990X},
  doi = {10.1103/PhysRevFluids.10.014003},
  urldate = {2025-02-24},
  langid = {english}
}

@article{Hu2002,
  title = {Evaporation of a {{Sessile Droplet}} on a {{Substrate}}},
  author = {Hu, Hua and Larson, Ronald G.},
  year = 2002,
  month = feb,
  journal = {The Journal of Physical Chemistry B},
  volume = {106},
  number = {6},
  pages = {1334--1344},
  issn = {1520-6106, 1520-5207},
  doi = {10.1021/jp0118322},
  urldate = {2019-07-05},
  langid = {english}
}

@article{Huber2015,
  title = {Soft Matter in Hard Confinement: Phase Transition Thermodynamics, Structure, Texture, Diffusion and Flow in Nanoporous Media},
  shorttitle = {Soft Matter in Hard Confinement},
  author = {Huber, Patrick},
  year = 2015,
  month = mar,
  journal = {Journal of Physics: Condensed Matter},
  volume = {27},
  number = {10},
  pages = {103102},
  issn = {0953-8984, 1361-648X},
  doi = {10.1088/0953-8984/27/10/103102},
  urldate = {2019-07-05},
  langid = {english}
}

@article{Hyvaluoma2006,
  title = {Simulation of Liquid Penetration in Paper},
  author = {Hyv{\"a}luoma, J. and Raiskinm{\"a}ki, P. and J{\"a}sberg, A. and Koponen, A. and Kataja, M. and Timonen, J.},
  year = 2006,
  month = mar,
  journal = {Physical Review E},
  volume = {73},
  number = {3},
  pages = {036705},
  issn = {1539-3755, 1550-2376},
  doi = {10.1103/PhysRevE.73.036705},
  urldate = {2024-11-15},
  copyright = {http://link.aps.org/licenses/aps-default-license},
  langid = {english}
}

@article{Jain2019,
  title = {Adsorption, {{Desorption}}, and {{Crystallization}} of {{Aqueous Solutions}} in {{Nanopores}}},
  author = {Jain, Piyush and Vincent, Olivier and Stroock, Abraham D.},
  year = 2019,
  month = mar,
  journal = {Langmuir},
  volume = {35},
  number = {11},
  pages = {3949--3962},
  issn = {0743-7463},
  doi = {10.1021/acs.langmuir.8b04307},
  urldate = {2019-06-20},
  copyright = {All rights reserved}
}

@article{Jensen2016,
  title = {Sap Flow and Sugar Transport in Plants},
  author = {Jensen, K. H. and {Berg-S{\o}rensen}, K. and Bruus, H. and Holbrook, N. M. and Liesche, J. and Schulz, A. and Zwieniecki, M. A. and Bohr, T.},
  year = 2016,
  month = sep,
  journal = {Reviews of Modern Physics},
  volume = {88},
  number = {3},
  pages = {035007},
  issn = {0034-6861, 1539-0756},
  doi = {10.1103/RevModPhys.88.035007},
  urldate = {2020-07-20},
  langid = {english}
}

@article{Joly2011,
  title = {Capillary Filling with Giant Liquid/Solid Slip: {{Dynamics}} of Water Uptake by Carbon Nanotubes},
  shorttitle = {Capillary Filling with Giant Liquid/Solid Slip},
  author = {Joly, Laurent},
  year = 2011,
  month = dec,
  journal = {The Journal of Chemical Physics},
  volume = {135},
  number = {21},
  pages = {214705},
  publisher = {American Institute of Physics},
  issn = {0021-9606},
  doi = {10.1063/1.3664622},
  urldate = {2020-12-28}
}

@article{Kelly2016,
  title = {Anomalous Liquid Imbibition at the Nanoscale: The Critical Role of Interfacial Deformations},
  shorttitle = {Anomalous Liquid Imbibition at the Nanoscale},
  author = {Kelly, Shaina and {Torres-Verd{\'i}n}, Carlos and Balhoff, Matthew T.},
  year = 2016,
  journal = {Nanoscale},
  volume = {8},
  number = {5},
  pages = {2751--2767},
  issn = {2040-3364, 2040-3372},
  doi = {10.1039/C5NR04462F},
  urldate = {2025-10-02},
  langid = {english}
}

@article{Khalil2020a,
  title = {Wettability-Defined Droplet Imbibition in Ceramic Mesopores},
  author = {Khalil, Adnan and Sch{\"a}fer, Felix and Postulka, Niels and Stanzel, Mathias and Biesalski, Markus and {Andrieu-Brunsen}, Annette},
  year = 2020,
  month = dec,
  journal = {Nanoscale},
  volume = {12},
  number = {47},
  pages = {24228--24236},
  publisher = {The Royal Society of Chemistry},
  issn = {2040-3372},
  doi = {10.1039/D0NR06650H},
  urldate = {2022-01-17},
  langid = {english}
}

@article{Kim2012,
  title = {Natural Drinking Strategies},
  author = {Kim, Wonjung and Bush, John W. M.},
  year = 2012,
  month = aug,
  journal = {Journal of Fluid Mechanics},
  volume = {705},
  pages = {7--25},
  issn = {0022-1120, 1469-7645},
  doi = {10.1017/jfm.2012.122},
  urldate = {2025-10-02},
  copyright = {https://www.cambridge.org/core/terms},
  langid = {english}
}

@article{Lei2016,
  title = {Nanoscale {{Capillary Flows}} in {{Alumina}}: {{Testing}} the {{Limits}} of {{Classical Theory}}},
  shorttitle = {Nanoscale {{Capillary Flows}} in {{Alumina}}},
  author = {Lei, Wenwen and McKenzie, David R.},
  year = 2016,
  month = jul,
  journal = {The Journal of Physical Chemistry Letters},
  volume = {7},
  number = {14},
  pages = {2647--2652},
  issn = {1948-7185, 1948-7185},
  doi = {10.1021/acs.jpclett.6b01021},
  urldate = {2025-09-29},
  langid = {english}
}

@article{Li2021,
  title = {Applications of Capillary Action in Drug Delivery},
  author = {Li, Xiaosi and Zhao, Yue and Zhao, Chao},
  year = 2021,
  month = jul,
  journal = {iScience},
  volume = {24},
  number = {7},
  pages = {102810},
  issn = {25890042},
  doi = {10.1016/j.isci.2021.102810},
  urldate = {2025-08-07},
  langid = {english}
}

@article{Liu2016,
  title = {Evaporation {{Limited Radial Capillary Penetration}} in {{Porous Media}}},
  author = {Liu, Mingchao and Wu, Jian and Gan, Yixiang and Hanaor, Dorian A. H. and Chen, C. Q.},
  year = 2016,
  month = sep,
  journal = {Langmuir},
  volume = {32},
  number = {38},
  pages = {9899--9904},
  issn = {0743-7463},
  doi = {10.1021/acs.langmuir.6b02404},
  urldate = {2019-10-19}
}

@article{Lucas1918,
  title = {{Ueber das Zeitgesetz des kapillaren Aufstiegs von Fl\"ussigkeiten}},
  author = {Lucas, Richard},
  year = 1918,
  month = jul,
  journal = {Kolloid-Zeitschrift},
  volume = {23},
  number = {1},
  pages = {15--22},
  issn = {1435-1536},
  doi = {10.1007/BF01461107},
  urldate = {2019-07-08},
  langid = {ngerman}
}

@article{Martinez2024,
  title = {Filling Fraction Measurement around a Drying Drop onto Nanoporous Silicon Using Digital Holographic Microscopy},
  author = {Mart{\'i}nez, M. F. and Sallese, M. D. and Psota, P. and Berli, C. L. A. and Urteaga, R. and Budini, N. and Monaldi, A. C.},
  year = 2024,
  month = feb,
  journal = {Journal of Applied Physics},
  volume = {135},
  number = {7},
  pages = {073102},
  issn = {0021-8979, 1089-7550},
  doi = {10.1063/5.0190518},
  urldate = {2024-03-27},
  langid = {english}
}

@book{Masoodi2017,
  title = {Wicking in Porous Materials: Traditional and Modern Modeling Approaches},
  shorttitle = {Wicking in Porous Materials},
  editor = {Masoodi, Reza and Pillai, Krishna M.},
  year = 2017,
  edition = {First issued in paperback},
  publisher = {CRS Press},
  address = {Boca Raton London New York},
  isbn = {978-1-138-07610-5 978-1-4398-7432-5},
  langid = {english}
}

@article{Mercuri2017,
  title = {Complex {{Filling Dynamics}} in {{Mesoporous Thin Films}}},
  author = {Mercuri, Magal{\'i} and Pierpauli, Karina and Bellino, Mart{\'i}n G. and Berli, Claudio L. A.},
  year = 2017,
  month = jan,
  journal = {Langmuir},
  volume = {33},
  number = {1},
  pages = {152--157},
  publisher = {American Chemical Society},
  issn = {0743-7463},
  doi = {10.1021/acs.langmuir.6b03987},
  urldate = {2020-12-28}
}

@book{Nobel2020,
  title = {Physicochemical and Environmental Plant Physiology},
  author = {Nobel, Park},
  year = 2020,
  publisher = {Elsevier},
  address = {Cambridge},
  isbn = {978-0-12-819146-0}
}

@article{Olanrewaju2018,
  title = {Capillary Microfluidics in Microchannels: From Microfluidic Networks to Capillaric Circuits},
  shorttitle = {Capillary Microfluidics in Microchannels},
  author = {Olanrewaju, Ayokunle and Beaugrand, Ma{\"i}wenn and Yafia, Mohamed and Juncker, David},
  year = 2018,
  journal = {Lab on a Chip},
  volume = {18},
  number = {16},
  pages = {2323--2347},
  issn = {1473-0197, 1473-0189},
  doi = {10.1039/C8LC00458G},
  urldate = {2025-09-30},
  langid = {english}
}

@article{Olivella2000,
  title = {Vapour {{Transport}} in {{Low Permeability Unsaturated Soils}} with {{Capillary Effects}}},
  author = {Olivella, S. and Gens, A.},
  year = 2000,
  month = aug,
  journal = {Transport in Porous Media},
  volume = {40},
  number = {2},
  pages = {219--241},
  issn = {0169-3913, 1573-1634},
  doi = {10.1023/A:1006749505937},
  urldate = {2025-09-30},
  copyright = {https://www.springernature.com/gp/researchers/text-and-data-mining},
  langid = {english}
}

@article{Page1993,
  title = {Pore-Space Correlations in Capillary Condensation in {{Vycor}}},
  author = {Page, J. H. and Liu, J. and Abeles, B. and Deckman, H. W. and Weitz, D. A.},
  year = 1993,
  month = aug,
  journal = {Physical Review Letters},
  volume = {71},
  number = {8},
  pages = {1216--1219},
  issn = {0031-9007},
  doi = {10.1103/PhysRevLett.71.1216},
  urldate = {2019-07-05},
  langid = {english}
}

@article{Pel2018,
  title = {A Simplified Model for the Combined Wicking and Evaporation of a {{NaCl}} Solution in Limestone},
  author = {Pel, L. and Pishkari, R. and Casti, M.},
  year = 2018,
  month = jun,
  journal = {Materials and Structures},
  volume = {51},
  number = {3},
  pages = {66},
  issn = {1359-5997, 1871-6873},
  doi = {10.1617/s11527-018-1187-y},
  urldate = {2025-10-02},
  langid = {english}
}

@article{Pizarro2024,
  title = {Ion--{{Fluid Transport-Control Feedback}} along {{Nanopore Networks}}},
  author = {Pizarro, Agustin D. and Berli, Claudio Luis Alberto and {Soler-Illia}, Galo J. A. A. and Bellino, Mart{\'i}n Gonzalo},
  year = 2024,
  month = jun,
  journal = {ACS Nano},
  volume = {18},
  number = {25},
  pages = {16199--16207},
  issn = {1936-0851, 1936-086X},
  doi = {10.1021/acsnano.4c01898},
  urldate = {2024-10-22},
  copyright = {https://doi.org/10.15223/policy-029},
  langid = {english}
}

@article{Sallese2020,
  title = {Optical Coherence Tomography Measurement of Capillary Filling in Porous Silicon},
  author = {Sallese, M. and Torga, J. and Morel, E. and Budini, N. and Urteaga, R.},
  year = 2020,
  month = jul,
  journal = {Journal of Applied Physics},
  volume = {128},
  number = {2},
  pages = {024701},
  publisher = {American Institute of Physics},
  issn = {0021-8979},
  doi = {10.1063/1.5145270},
  urldate = {2020-07-15}
}

@article{Seker2008,
  title = {Kinetics of Capillary Wetting in Nanoporous Films in the Presence of Surface Evaporation},
  author = {Seker, Erkin and Begley, Matthew R. and Reed, Michael L. and Utz, Marcel},
  year = 2008,
  month = jan,
  journal = {Applied Physics Letters},
  volume = {92},
  number = {1},
  pages = {013128},
  publisher = {American Institute of Physics},
  issn = {0003-6951},
  doi = {10.1063/1.2831007},
  urldate = {2021-07-28}
}

@article{Shokri-Kuehni2017,
  title = {New Insights into Saline Water Evaporation from Porous Media: {{Complex}} Interaction between Evaporation Rates, Precipitation, and Surface Temperature},
  shorttitle = {New Insights into Saline Water Evaporation from Porous Media},
  author = {Shokri-Kuehni, Salom{\'e} M. S. and Vetter, Thomas and Webb, Colin and Shokri, Nima},
  year = 2017,
  month = jun,
  journal = {Geophysical Research Letters},
  volume = {44},
  number = {11},
  pages = {5504--5510},
  issn = {0094-8276, 1944-8007},
  doi = {10.1002/2017GL073337},
  urldate = {2025-09-30},
  copyright = {http://onlinelibrary.wiley.com/termsAndConditions\#vor},
  langid = {english}
}

@article{Siefert2025,
  title = {Elastocapillary Sequential Fluid Capture in Hummingbird-Inspired Grooved Sheets},
  author = {Si{\'e}fert, Emmanuel and Scheid, Benoit and Brau, Fabian and Cappello, Jean},
  year = 2025,
  month = may,
  journal = {Nature Communications},
  volume = {16},
  number = {1},
  pages = {4913},
  issn = {2041-1723},
  doi = {10.1038/s41467-025-60203-8},
  urldate = {2025-09-30},
  langid = {english}
}

@article{Thommes2015,
  title = {Physisorption of Gases, with Special Reference to the Evaluation of Surface Area and Pore Size Distribution ({{IUPAC Technical Report}})},
  author = {Thommes, Matthias and Kaneko, Katsumi and Neimark, Alexander V. and Olivier, James P. and {Rodriguez-Reinoso}, Francisco and Rouquerol, Jean and Sing, Kenneth S.W.},
  year = 2015,
  month = oct,
  journal = {Pure and Applied Chemistry},
  volume = {87},
  number = {9-10},
  pages = {1051--1069},
  issn = {1365-3075, 0033-4545},
  doi = {10.1515/pac-2014-1117},
  urldate = {2023-05-11},
  langid = {english}
}

@article{Todorova2022,
  title = {Improvement of Barrier Properties for Packaging Applications},
  author = {Todorova, Dimitrina and Yavorov, Nikolay and Lasheva, Veska},
  year = 2022,
  month = jun,
  journal = {Sustainable Chemistry and Pharmacy},
  volume = {27},
  pages = {100685},
  issn = {23525541},
  doi = {10.1016/j.scp.2022.100685},
  urldate = {2025-09-30},
  langid = {english}
}

@article{Urteaga2019,
  title = {Spontaneous Water Adsorption-Desorption Oscillations in Mesoporous Thin Films},
  author = {Urteaga, Ra{\'u}l and Mercuri, Magal{\'i} and Gimenez, Roc{\'i}o and Bellino, Martin G. and Berli, Claudio L. A.},
  year = 2019,
  month = mar,
  journal = {Journal of Colloid and Interface Science},
  volume = {537},
  pages = {407--413},
  issn = {0021-9797},
  doi = {10.1016/j.jcis.2018.11.055},
  urldate = {2019-07-06}
}

@article{vanHonschoten2010,
  title = {Capillarity at the Nanoscale},
  author = {{van Honschoten}, Joost W. and Brunets, Nataliya and Tas, Niels R.},
  year = 2010,
  journal = {Chemical Society Reviews},
  volume = {39},
  number = {3},
  pages = {1096},
  issn = {0306-0012, 1460-4744},
  doi = {10.1039/b909101g},
  urldate = {2019-07-05},
  langid = {english}
}

@article{Venditti2021,
  title = {Chromatographic {{Effects}} in {{Inkjet Printing}}},
  author = {Venditti, Gianmarco and Murali, Vignesh and Darhuber, Anton A.},
  year = 2021,
  month = oct,
  journal = {Langmuir},
  volume = {37},
  number = {40},
  pages = {11726--11736},
  issn = {0743-7463, 1520-5827},
  doi = {10.1021/acs.langmuir.1c01624},
  urldate = {2022-05-31},
  langid = {english}
}

@article{Vincent2014,
  title = {Drying by {{Cavitation}} and {{Poroelastic Relaxations}} in {{Porous Media}} with {{Macroscopic Pores Connected}} by {{Nanoscale Throats}}},
  author = {Vincent, Olivier and Sessoms, David A. and Huber, Erik J. and Guioth, Jules and Stroock, Abraham D.},
  year = 2014,
  month = sep,
  journal = {Physical Review Letters},
  volume = {113},
  number = {13},
  pages = {134501},
  doi = {10.1103/PhysRevLett.113.134501},
  urldate = {2018-11-27},
  copyright = {All rights reserved}
}

@article{Vincent2016,
  title = {Capillarity-Driven Flows at the Continuum Limit},
  author = {Vincent, Olivier and Szenicer, Alexandre and Stroock, Abraham D.},
  year = 2016,
  month = aug,
  journal = {Soft Matter},
  volume = {12},
  number = {31},
  pages = {6656--6661},
  issn = {1744-6848},
  doi = {10.1039/C6SM00733C},
  urldate = {2018-11-27},
  copyright = {All rights reserved},
  langid = {english}
}

@article{Vincent2017,
  title = {Imbibition {{Triggered}} by {{Capillary Condensation}} in {{Nanopores}}},
  author = {Vincent, Olivier and Marguet, Bastien and Stroock, Abraham D.},
  year = 2017,
  month = feb,
  journal = {Langmuir},
  volume = {33},
  number = {7},
  pages = {1655--1661},
  issn = {0743-7463},
  doi = {10.1021/acs.langmuir.6b04534},
  urldate = {2018-11-27},
  copyright = {All rights reserved}
}

@article{Vincent2024,
  title = {Tunable Transport in Bidisperse Porous Materials with Vascular Structure},
  author = {Vincent, Olivier and Tassin, Th{\'e}o and Huber, Erik J. and Stroock, Abraham D.},
  year = 2024,
  month = jun,
  journal = {Physical Review Fluids},
  volume = {9},
  number = {6},
  pages = {064202},
  issn = {2469-990X},
  doi = {10.1103/PhysRevFluids.9.064202},
  urldate = {2024-07-12},
  copyright = {All rights reserved},
  langid = {english}
}

@article{Washburn1921,
  title = {The {{Dynamics}} of {{Capillary Flow}}},
  author = {Washburn, Edward W.},
  year = 1921,
  month = mar,
  journal = {Physical Review},
  volume = {17},
  number = {3},
  pages = {273--283},
  issn = {0031-899X},
  doi = {10.1103/PhysRev.17.273},
  urldate = {2019-07-05},
  langid = {english}
}

@article{Weickgenannt2011,
  title = {Nonisothermal Drop Impact and Evaporation on Polymer Nanofiber Mats},
  author = {Weickgenannt, Christina M. and Zhang, Yiyun and Lembach, Andreas N. and Roisman, Ilia V. and {Gambaryan-Roisman}, Tatiana and Yarin, Alexander L. and Tropea, Cameron},
  year = 2011,
  month = mar,
  journal = {Physical Review E},
  volume = {83},
  number = {3},
  pages = {036305},
  issn = {1539-3755, 1550-2376},
  doi = {10.1103/PhysRevE.83.036305},
  urldate = {2025-10-02},
  copyright = {http://link.aps.org/licenses/aps-default-license},
  langid = {english}
}

@article{Zhong2018,
  title = {Capillary {{Condensation}} in 8 Nm {{Deep Channels}}},
  author = {Zhong, Junjie and Riordon, Jason and Zandavi, Seyed Hadi and Xu, Yi and Persad, Aaron H. and Mostowfi, Farshid and Sinton, David},
  year = 2018,
  month = feb,
  journal = {The Journal of Physical Chemistry Letters},
  volume = {9},
  number = {3},
  pages = {497--503},
  publisher = {American Chemical Society},
  doi = {10.1021/acs.jpclett.7b03003},
  urldate = {2021-06-06}
}

@article{Zhou2019,
  ids = {Zhou2019a,Zhou2019b},
  title = {Wetting Enhanced by Water Adsorption in Hygroscopic Plantlike Materials},
  author = {Zhou, Meng and Car{\'e}, Sabine and King, Andrew and {Courtier-Murias}, Denis and Rodts, St{\'e}phane and Gerber, Ga{\'e}tan and Aimedieu, Patrick and Bonnet, Marie and Bornert, Michel and Coussot, Philippe},
  year = 2019,
  month = dec,
  journal = {Phys. Rev. Research},
  volume = {1},
  number = {3},
  pages = {033190},
  publisher = {[object Object]},
  doi = {10.1103/PhysRevResearch.1.033190},
  numpages = {[object Object]}
}

\end{document}